\documentclass[11pt]{article}

\usepackage{graphicx}

\usepackage{graphicx,psfrag}

\def \s {\sigma}

\def \i {{\hbox {i}}}

\def \l {\lambda}

\def \vep {\varepsilon}

\def \a {\alpha}

\def \T {{\hbox {\ Tr}}}

\def \P {\Psi}

\def \b {\bar}

\def \t {\tilde}

\def \h {\hat}

\def \E {{\bf E}}

\def \i {{\hbox{ i}}}

\def \vep {\varepsilon}

\def \th {\theta}

\begin{document}
\title{Estimates for moments of random matrices with Gaussian
elements}

\author{O. Khorunzhiy}


\maketitle

\begin{abstract}

We describe an elementary method to get non-asymptotic estimates for the
moments of Hermitian random matrices 
whose elements are Gaussian independent
random variables. 
We derive a system of recurrent relations for the moments and the covariance
terms and develop a triangular scheme to prove the recurrent estimates.
The estimates we obtain are asymptotically exact
in the sense that they give exact expressions
for the first terms of $1/N$-expansions of the moments and covariance terms.

As the basic example, we consider 
the Gaussian Unitary Ensemble of random \mbox{matrices} (GUE).
Immediate applications include the Gaussian {Ortho\-gonal} Ensemble  and the ensemble of Gaussian
anti-symmetric Hermitian matrices. 
Finally we apply our method to the ensemble of $N\times N$ Gaussian
Hermitian random matrices 
$H^{(N,b)}$ 
whose elements are zero outside of the band of width $b$.
The other elements are taken from GUE;  the matrix obtained  
 is renormalized by $ b^{-1/2}$.
We derive the estimates for the moments of $H^{(N,b)}$ and prove that the spectral norm $\Vert H^{(N,b)}\Vert$ 
remains bounded in the limit $N,b\to\infty$ when $ (\log N)^{3/2}/b \to 0 $. 
\end{abstract}

\section{Introduction}

The moments of $N\times N$ Hermitian random matrices $H_N$
are given by expression
$$
M_{k}^{(N)} = \E\left\{ {1\over N} \T \, (H_N)^{k}\right\},
$$ 
where $\E\{ \cdot\} $ denotes the corresponding mathematical expectation.
The asymptotic  behavior of $M_{k}^{(N)}$ in the limit $N\to\infty$
is the source of numerous studies and vast list of  publications. 
One can  observe  three main directions of researches; 
we list and mark them with the
references that are earliest in the field up to our knowledge.

The first group of results is related with the limiting transition $N\to\infty$ 
when  the numbers $k$ are fixed. 
In this case the limiting values 
of $M_{k}^{(N)}$, if they exist,  determine the
moments $m_k$ of the limiting spectral measure $\sigma$ of the ensemble $\{H_N\}$. This
problem was considered first by E. Wigner 
\cite{W}.

Another asymptotic regime, when $k$ goes to infinity
at the same time as $N$ does, is more informative
and can be considered in two particular cases.
In the first one  $k$ grows slowly and $1\ll k\ll N^\gamma$ for any $\gamma>0$.
In particular, if  $k$ is of the order $\log N$ or greater, 
the maximal eigenvalue of $H_N$ dominates in 
 the asymptotic behavior of $M_{2k}^{(N)}$. 
 Then the exponential estimates of
$M_{2k}^{(N)}$ provide the asymptotic bounds for the probability of deviations
of the spectral norm $\Vert H_N\Vert$. 
This observation due to U. Grenander has originated a series
of deep results started by S. Geman
\cite{BY,FK,G}.

The second asymptotic regime is related to the limit when $k = O(N^\gamma)$
with  $\gamma>0$.
The main subject here is to determine the critical 
exponent 
$\tilde \gamma$ such that the same estimates for $M_{2k}^{(N)}$ 
as in the previous case 
 remain valid for all $\gamma\le \tilde \gamma$ and fail otherwise \cite{S}. This allows one to conclude about the
order of the mean distance between eigenvalues 
 at the border of the support of the limiting spectral density $d\sigma$
\cite{Br,TW}. 

In present article we describe a method to get the estimates of
$M_{2k}^{(N)}$ that are valid for all values of $N$ and $k$ such that 
$k\le C N^{\tilde \gamma}$ with some constant $C$. The estimates of this type are called non-asymptotic. However, they remain valid in the limit $N\to\infty$ and in this case they belong to the second asymptotic regime.

As the basic example, we consider the Gaussian Unitary
(Invariant)
Ensemble of random matrices that is usually abbreviated as GUE. 
  In section 2 we describe our method and prove the main results
for GUE.
Immediate applications of our method include the Gaussian Orthogonal (Invariant)
Ensemble of random matrices (GOE) and the 
Gaussian anti-symmetric (or skew-symmetric)  
Hermitian random matrices with
independent elements. 
Detailed description of these ensembles is given in monograph
\cite{M}. In section 3 we present  the non-asymptotic estimates 
for the corresponding moments.
 
Our approach is elementary. We use the integration by parts formula and generating functions technique only. We do not employ such a powerful 
methods like the orthogonal polynomials technique
commonly applied to unitary and orthogonally  invariant  random matrix ensembles.
 This allows us to consider more general ensembles of random matrices  than GUE and GOE.
One of the possible developments is given by the study of
the ensemble of Hermitian  band random matrices $H^{(N,b)}$. The matrix elements 
of $H^{(N,b)}$ 
 within the band of the width
$b$ along the principal diagonal coincide with those of GUE.
Outside of this band they are equal to zero; the matrix obtained is normalized by 
$b^{-1/2}$.
In section 4 we prove
non-asymptotic estimates for the  moments of $H^{(N,b)}$. 
These estimates allow us to
conclude about the asymptotic behavior 
of the spectral norm $\Vert H^{(N,b)}\Vert$ in the limit $b,N\to\infty$. 

In section 5 we collect  auxiliary computations and formulas used.

\subsection{GUE, recurrent relations and semi-circle law}

GUE is determined by the probability distribution 
over the set of Hermitian matrices $\{ H_N\}$
with the density proportional to
$$
 \exp \{ -2N \T H_N^2\}.
\eqno (1.1)
$$
The odd moments of $H_N$ are zero and the even ones $M_{2k}^{(N)}$ verify the following 
remarkable recurrent
relation discovered by 
 Harer and Don Zagier \cite{H-Z} 
$$
M_{2k}^{(N)} = {2k-1\over 2k+2} \ M_{2k-2}^{(N)} +
{2k-1\over 2k+2} \cdot {2k-3\over 2k}\cdot  {k(k-1)\over 4N^2}\ 
M_{2k-4}^{(N)},
\eqno (1.2)
$$
where $M_{0}^{(N)} =1$ and $M_2^{(N)}=1/4$.
 It follows from (1.2) that  the moments $M_{2k}^{(N)}, k=0,1,...$ converge as $N\to\infty$ to 
the limiting $m_k$ determined by relations
$$
m_k = {2k-1\over 2k+2} \ m_{k-1}, \quad m_0=1.
\eqno (1.3)
$$
The limiting  moments $\{m_k, k\ge 0\}$ are proportional to the Catalan numbers
$C_k$:
$$
m_k = {1\over 4^k}{1\over  (k+1)} {2k \choose k} = {1\over 4^k} C_k
\eqno (1.4)
$$
and therefore verify the following recurrent relation
$$
m_k = {\displaystyle 1\over \displaystyle 4} \ \sum_{j=0}^{k-1} \ m_{k-1-j}\
m_j, \quad k= 1,2, \dots
\eqno(1.5)
$$
with obvious initial condition $m_0=1$.

In random matrix theory, equality (1.5) 
was observed for the first time by E. Wigner
\cite{W}.
Relation (1.5) implies that 
the generating function of the moments $m_k$ 
 $$
f(\tau) = \sum_{k= 0}^\infty  \ m_k\cdot  \tau^k
$$ 
verifies quadratic equation
$   \tau f^2(\tau)-4f(\tau) +4 =0$
and is given by 
$$
 f(\tau) = {1 - \sqrt{1-\tau}\over \tau/2}   .
\eqno (1.6)
$$
Using (1.6), Wigner has shown that the measure $\sigma_w$ determined by the moments $m_k = \int \lambda ^{2k} \ d \sigma_{w}(\lambda)$
has the density  of the semicircle form
$$
\sigma_w'(\lambda) = {2\over\pi} \cases{ \sqrt{1 - \lambda^2}, & if $\vert
\lambda \vert \le 1$, \cr
0, & if $\vert \lambda \vert >1.$\cr}
\eqno (1.7)
$$
The statement that the moments $M_{l}^{(N)}$ converge to $m_k$ for $l=2k$ and to $0$ for $l=2k+1$ is known as the Wigner semicircle law. 

In present paper we show that the generating function $f(\tau)$ 
together with its derivatives
represents a very convenient tool when estimating the moments $M_{2k}^{(N)}$.
Everywhere below, we use denotation   $[\cdot]_k$ for the $k$-th coefficient of the corresponding
development, so   $[f(\tau)]_k= m_k$.

\subsection{Estimates for the moments of GUE}

Using relations (1.2) and (1.3), one can easily prove by induction  the  estimate
$$
M_{2k}^{(N)} \le \left( 1 + {k^2\over 8N^2}\right)^{2k}\,  m_{k}.
\eqno (1.8)
$$
Indeed, let us assume inequalities $M_{2l}^{(N)}\le (1+l^2/(gN^2))^{2l} m_l$ with 
some $g>0$  to hold
for all values of $l$ such that $1\le l\le k-1$. 
Let us show that this is also true for $l=k$ provided $g\le 8$.

Regarding the right-hand side of (1.2) and
replacing $M_{2k-2}^{(N)}$ and $M_{2k-4}^{(N)}$ by corresponding estimates
with $l=k-1$ and $l=k-2$, respectively, we  bound the right-hand side of (1.2) by
the sum of $$
 {2k-1\over 2k+2} \ \left( 1+ {(k-1)^2\over gN^2}\right)^{2k-2} m_{k-1} 
 = \left( 1+ {(k-1)^2\over gN^2}\right)^{2k-2} m_{k}
 $$
 and
 $$
 {k(k-1)\over 4N^2} \left(1 + {(k-2)^2\over gN^2}\right)^{2k-4} m_{k}.
$$ 
Here we have used identity (1.3).
Comparing the expression obtained with 
the right-hand side of (1.8), we see that  the following inequality
$$
 \left( 1+ {(k-1)^2\over gN^2}\right)^{2} +
  {k(k-1)\over 4N^2}\le 
\left(1 + {k^2\over gN^2}\right)^{4} 
$$
is sufficient for (1.8) to be true.  Expanding the powers, we see that 
the condition $g\le 8$ is sufficient to have (1.8) valid for all values of $k$ and $N$.

Estimates (1.8) are valid for all
values of $k$ and $N$ without any restriction.
They allow one to estimate the probability of deviations of the largest
eigenvalue of $H_N$ (see, for example \cite{L,L2} and references therein).
Then one can study the asymptotic behavior
of the maximal eigenvalues and also conclude about
spectral scales 
at the borders of the support of $\s_w'$ (see \cite{S}).

It should be noted that relations (1.2) are obtained 
in \cite{H-Z} with the help of the orthogonal polynomials
technique (see \cite{HT} and \cite{L2} for the simplified derivation).  
There are several
 more random matrix ensembles (see \cite{L2} for the references)
whose moments verify recurrent relations of the type (1.2).
But relations of the type (1.2) are rather exceptional than typical.
Even in the case of GOE, it is not known
 whether relations of the type (1.2) exist. 
As a result, 
 no simple derivation of the estimates of the form (1.8) for GOE 
has been reported.

We develop one more approach to prove  non-asymptotic estimates of the type (1.8).
Instead of relations (1.2), we use the system of recurrent relations
(1.5) that  is of more
general character than (1.2). 
Regarding various random matrix ensembles,
one can observe that the limiting moments verify either (1.5) by itself or 
 one or another system recurrent relations generalizing (1.5)
(see for instance, section 5 of  \cite{BK}, where the first elements of the present approach were presented).

We derive a system of recurrent relations for the moments $M_{2k}^{(N)}$
that have (1.5) as the limiting form. 
These relations for $M_{2k}^{(N)}$ involve  corresponding covariance terms.
Using the generation functions technique, we find the form of estimates
and use the
triangle scheme of the recurrent estimates to prove
the bounds for moments and covariance terms. The final result can be written as
$$
M_{2k}^{(N)} \le \left( 1 + \alpha {k^3\over N^2}\right) m_k
\eqno (1.9)
$$  
with some $\alpha >1/12$.
The estimates obtained are valid in the domain  $k^3\le \chi N^2$ 
with some constant $\chi$, i.e. not for all values of $k$ and $N$, as (1.8) does.
But in this region our estimates are more precise than those of (1.8).
If $k^3 \ll N^2$, our estimates provide exact expressions for $1/N$-corrections
for the moments $M_{2k}^{(N)}$.

\subsection{Band random matrices and the semi-circle law}

Hermitian band random matrices $H^{(N,b)}$ can be  obtained from GUE matrices by erasing all elements outside of the band of width $b$ along the
principal diagonal and by renormalizing the matrix obtained  by the factor $b^{-1/2}$. It appears
that the limiting values of the moments 
$$
M_{2k}^{(N,b)}= \E \left\{{1\over N}   \T\left(H^{(N,b)}\right) ^{2k}\right\}
$$
crucially depend of the ratio between $b$ and $N$ when $N\to\infty$ (see \cite{CG,KLH,MPK}).

If $b/N \to 1$ as $N\to \infty$, then $M_{2k}^{(N,b)} \to m_k$
and  the semicircle law is valid in this case.
If  $b/N \to c $ and  $0<c<1$, then the limiting values of $M_{2k}^{(N,b)}$ differ from $m_k$.
 Finally, if $1\ll b\ll N$, then the semicircle law is valid again.

The last asymptotic regime of (relatively) narrow band width
attracts a special interest of researchers. In this case the
spectral properties of  band random matrices 
exhibit a transition  from one type to another.
The first one is characterized by GUE matrices and the second
is given by
spectral properties of Jacobi random matrices, i.e. the discrete
analog of the random  Schr\"odinger operator with $b=3$ (see \cite{CMI,FM} for the results
and references). It is shown that the value $b' = \sqrt N$ is critical
with respect to this transition \cite{CMI,FM,KK}.

In present paper we derive the estimates for  $M^{(N,b)}_{2k}$
that have the same form as the estimates for GUE with $N$ replaced by $b$.
This can be viewed as an evidence to the fact that the
asymptotic behavior of the eigenvalues of $H^{(N,b)}$ at the border of the 
semi-circle density is similar to that of matrices of the size $b\times b$. 
The estimates we obtain show that the value  $b'= \sqrt N$ does not play any
particular role with respect to the asymptotic behavior of the  spectral norm 
$\Vert H^{(N,b)}\Vert$.
We show that if $b\gg (\log N)^{3/2}$, then the spectral norm
converges  with probability 1 when $N\to\infty$ to the edge of the corresponding 
semicircle
density. Up to our knowledge, this is the first
result on the upper bound of the spectral norm 
of band random matrices.

\section{Gaussian Hermitian Ensembles}

Let us consider the family of complex random variables 
$$
h_{xy} = \cases{ V_{xy} + \i W_{xy}, & if $x\le y$,\cr
  V_{yx} - \i W_{yx}, & if $x>y$,\cr}
\eqno (2.1)
$$
where $\{ V_{xy}, W_{xy}, \ 1\le x\le y\le N\}$ are real jointly independent
random variables that have normal (Gaussian) distribution with the properties
$$
\E V_{xy} = \E W_{xy} = 0,
\eqno (2.2a)
$$
and 
$$
\E V_{xy}^2 = ({1 + \delta_{xy}} ){1+ \eta \over 8}\, ,
\quad 
\E W_{xy}^2 = ({1 - \delta_{xy}} ) {1- \eta \over 8}\, ,
\eqno (2.2b)
$$
where $\delta_{xy}$ is the Kronecker $\delta$-symbol and  $\eta \in [-1,1]$.
Then we obtain the family of Gaussian ensembles of
$N\times N$ Hermitian random matrices of the form
$$
(H_N^{(\eta)})_{xy} = {1\over \sqrt N} \, h_{xy}, \quad x,y =1, \dots, N
\eqno (2.3)
$$
that generalizes the  Gaussian Unitary Ensemble (1.1). Indeed, it is easy to see that  
$\{H^{(0)}_N\}$ coincides
with the GUE, while $\{H^{(1)}\}$ and $\{H^{(-1)}\}$ reproduce
the GOE and Hermitian skew-symmetric Gaussian matrices. In \cite{M},
the last ensemble is referred to as the Hermitian anti-symmetric one; 
below we follow this
terminology.  The present section is devoted to the results for GUE and their proofs. 
Two other ensembles will be considered
in the section 3.

\subsection{Main results for GUE and the scheme of the proof}

Let us consider the moments $M_{2k}^{(N)}$ of GUE matrices. We prove a little more precise  estimate than (1.9).

\vskip 0.2cm
\noindent {\bf Theorem 2.1}

\noindent {\it Given any constant $\alpha  > 1/12$, there exists $\chi>0$
such that the estimate
$$
M_{2k}^{(N)} \le   \left( 1 + \alpha { k(k^2-1)\over N^{2}} \right) m_k
\eqno (2.4)
$$
holds for all values of $k,N$ under condition that $k^3/N^2 \le \chi$.
}

\vskip 0.3 cm
\noindent {\it Remark.} 
 Using relation (1.2), one can prove  (2.4)
under condition that 
$$
\alpha > {1\over 12- \chi}.
\eqno (2.5)
$$
This relation shows that  Theorem 2.1 gives the correct lower bound for
$\alpha$.  In our proof we get relations between $\chi $ and $\alpha$ more
complicated than (2.5), but they are of the same character as (2.5).
It follows from (2.5) that the closer $\alpha$ to $1/12$ is, the
smaller
$\chi$ has to be chosen and vice versa. Indeed, the following proposition shows that the estimate (2.4)
is asymptotically exact.

\vskip 0.5cm

\noindent {\bf Theorem 2.2}

\noindent {\it 
Given $k$ fixed,
 the following asymptotic expansion holds,
$$
M_{2k}^{(N)} = m_k + {1\over N^2} m_k^{(2)} + O(N^{-4}),
\quad {\hbox{as
}}N\to\infty,
\eqno (2.6a)
$$
where
$$
m_k^{(2)} = { k(k-1)(k+1)\over 12} \ m_k, \quad k\ge 1.
\eqno (2.6b)
$$
If $k\to \infty$ and $\t\chi = k^3/N^2 \to 0$, then 
relation (2.6a) remains true with $O(N^{-4})$ replaced by $o(\t\chi)$.}

\vskip 0.5cm

\noindent {\it Remark.} 
It follows from  (1.2) that the sequence $\{m_k^{(2)}, k\ge 1\}$
is determined by recurrent relation
$$
m_k^{(2)} = {2k-1\over 2k+2} \cdot m_{k-1}^{(2)} +
{k(k-1)\over 4} \cdot m_k,\quad k= 1,2,\dots
$$
with obvious initial condition $m^{(2)}_0 = 0$.
It is easy to check that (2.6b) is in complete agreement with this recurrent relation
for $m_k^{(2)}$.

\vskip 0.5cm

Let us explain the role of recurrent relations (1.5) in the proof of Theorem 2.1. To do this, 
let us consider the
normalized trace $L_a ={1\over  N} \, \T \, H^a$ 
$$
 \E \{ L_a \} = {1\over N} \sum_{x,s=1}^N
\E \{ H_{xs} H^{a-1}_{sx} \}
$$
and compute the last mathematical expectation.
Here and below we omit subscripts and superscripts $N$ when no confusion can arise.
 Applying the integration by parts formula
(see section 5 for details), we obtain equality
$$
\E \{ L_a \} = 
{1\over 4} \sum_{j=0}^{a-2} \E \{ L_{a-2-j} L_{j} \}.
\eqno (2.7)
$$
Introducing the centered random variables $L^o_j = L_j - \E L_j$, we can write that 
$$
\E \left\{ L_{a_1} L_{a_2} \right\} = 
\E\{ L_{a_1}\} \, \E \{L_{a_2}\} +\E \{ L^o_{a_1} L^o_{a_2}\}.
$$
Taking into account that 
$ \E L_{2k+1} = 0$, we deduce from (2.7) 
relation 
$$
M_{2k}^{(N)} = {1\over 4}\sum_{j=0}^{k-1}
M_{2k-2-2j}^{(N)} \, M_{2j}^{(N)} +
{1\over 4} \ D_{2k-2}^{(2;N)},
\eqno (2.8)
$$
where we denoted
$$
D_{2k-2}^{(2;N)}= \sum_{a_1+a_2 =2k-2} \E \{ L^o_{a_1} L^o_{a_2}\}.
$$
Obviously, the last summation runs over $a_i>0$. Comparing (2.8) with (1.5), we see
that the problem is to  estimate the covariance terms $D^{(2)}$. Here and below we omit superscripts $N$ when no confusion can arise.

In what follows, we prove that under conditions of Theorem 2.1,
$$
\vert D^{(2;N)}_{2k}\vert \le {ck\over N^2}, 
\eqno (2.9)
$$
with some constant $c$.
Inequality (2.9) represents the main technical result of this paper. It is proved in the next subsection.
With (2.9) in hands,  we can use relation (2.8) to show that 
(2.4) holds.

Now let us explain  the use of the generating function $f(\tau)$ (1.6).
Regarding the right-hand side of (2.4), one can  observe that the third derivative
of $f(\tau)$ could be useful in computations because of the equality
$[f'''(\tau)]_k = (k+3)(k+2)(k+1)m_{k+3}$. 
 Indeed, more 
accurate computations (see identity (5.12) of section 5) 
show that the function
$$
f(\tau) +{A\over N^2} {\tau^2\over (1-\tau)^{5/2}} = \Phi_N(\tau) \quad \hbox{with}\quad  A = {3\alpha\over 4}
\eqno (2.10a)
$$
is a very good candidate to generate the estimating expressions. This is not by a simple coincidence or an artificial choice. Later we will see that the form of $\Phi_N(\tau)$
is in certain sense  optimal. It is dictated by the iteration scheme we use to get 
$1/N$-corrections for the moments and covariance terms (see subsection 2.5, the proof of Theorem 2.2).

Let us now show how (2.9) implies the estimate
$$
M_{2k}^{(N)} \le \left[ \Phi_N(\tau) \right]_k.
\eqno (2.10b)
$$
Assuming that this estimates and (2.9) are valid
for all the terms of the right-hand side of (2.8), we can estimate it with the help of
inequalities
$$
{1\over 4} \sum_{j=0}^{k-1}M_{2k-2-2j} M_{2j} + 
{1\over 4} \vert D^{(2)}_{2k-2} \vert \le
{1\over 4}\left[  \Phi^2_N(\tau) \right]_{k-1} +
{c\over 4N^2}\left[ {1 \over (1-\tau)^2}\right]_{k-2}.
$$
Denoting by $\Theta(k;N)$ the terms of the order $O(N^{-4})$, we can write that
$$
\left[ { \tau\over 4} \Phi^2_N(\tau) \right]_k = 
\left[{\tau f^2(\tau)\over 4} 
 +  {\tau^3 f(\tau)\over 2} {A\over N^2(1-\tau)^{5/2}} \right]_k  + \Theta(k;N).
$$
Rewriting (1.6) and quadratic equation for $f(\tau)$ in convenient  forms
$$
{\tau f^2(\tau)\over 4} = f(\tau)-1 \quad {\hbox{and}} \quad 
{\tau f(\tau)\over 2} = 1 - \sqrt{1-\tau},
\eqno (2.11)
$$
we transform the expression in the  brackets:
$$
\left[ f(\tau) + {A\over N^2} {\tau^2\over (1-\tau)^{5/2}} - 
{A\over N^2} {\tau^2\over (1-\tau)^2}\right]_k
= \left[ \Phi_N(\tau)\right]_k  - 
{A\over N^2} \left[{\tau^2\over (1-\tau)^2}\right]_k.
$$
Remembering that $[\Phi_N(\tau)]_k$ reproduces the expression to estimate  
$M_{2k}^{(N)} $, we conclude  that (2.10) is valid provided
 $$
{A\over N^2} \left[ {\tau^2\over (1-\tau)^{2}}\right]_k \ge  {c\over 4N^2} 
\left[{\tau^2\over (1-\tau)^2} \right]_k = {c(k-1)\over 4N^2}.
\eqno (2.12)
$$
This requires inequality $A \ge c/4$.

The final comment is related to the role of the
terms $\Theta(k;N)$. 
They are of the form 
$$
\Theta(k;N)= {A^2\over 4 N^4} \left[{\tau^5\over (1-\tau)^5}\right]_k \le {A^2 k^4\over N^4}.
$$
If one wants these terms not to violate inequality (2.12) involving terms of the form 
$k/N^2$, one has to set the ratio
$k^3/N^2 =\tilde  \chi$ sufficiently small. This explains the last condition of Theorem 2.1.

It should be noted that the same comments  concern the proof of the estimate of covariance terms (2.9), where the recurrent relations, generating functions and 
terms of the type $\tilde \chi$ appear. In the proofs, we constantly use relations (2.11).


\subsection{Main technical result}

In this subsection we prove the estimates of the covariance terms of the type $D^{(2)}_{2k}= \sum \E \{ L_{a_1}^o L^o_{a_2}\}$.
The main idea is that these terms are determined by a system of recurrent 
relations similar to (2.8). 
These relations involve the terms of more complicated structure than $D^{(2)}$.
The variables we study are defined as
\vskip 0.1cm
$$
D^{(q)}_{2k} = \sum_{ a_1+ \dots +a_q= 2k} D^{(q)}_{a_1,\dots,a_q} = 
{\sum_{a_1+ \dots + a_q=2k}} \E \left\{ L^o_{a_1} L^o_{a_2} \cdots
L^o_{a_q}\right\}, \quad q\ge 2.
$$
Here and everywhere below, we assume that the summation runs over all positive integers $a_i>0$.
\vskip 0.3cm
\noindent Our  main technical result is given by the following statement.

\vskip 0.5cm
\noindent {\bf Proposition 2.1.}

\noindent {\it Given 
$A>1/16$, there exists  $\chi>0$
such that 
 estimate (2.10) holds for all values of $1\le k\le k_0$, where $k_0$ verifies condition
$$
{ k_0^3\over N^2 } \le \chi\ .
\eqno (2.13)
$$
Also there exists $C$
$$
{1\over 24} < C < \max\{ {2A\over 3}, 4!\}
\eqno (2.14)
$$ 
such that inequalities 
$$
\vert D_{2k}^{(2s)}\vert  \le  C {(3s)!\over  N^{2s}} 
\left[{\tau\over  (1-\tau)^{2s}}\right]_{k}, 
\eqno (2.15a)
$$
and
$$
\vert D_{2k}^{(2s+1)}\vert  \le C^{} {(3s+3)! \over  N^{2s+2}} 
\left[{\tau \over (1-\tau)^{2s+ 5/2}} \right]_{k},
\eqno (2.15b)
$$
are true for 
all $k,s$ such that 
$$
2k+q\le 2 k_0
\eqno (2.16)
$$
with $q=2s$ and $q=2s+1$, respectively.
}

\vskip 0.5cm
\noindent {\it Remark.} The  form of estimates  (2.15) 
is dictated by the structure of the
recurrent relations
we derive below. 
The bounds for the constants $A$ and $C$ and of the form  factorial terms of (2.15)
are explained in subsection 2.4.

\vskip 0.5cm
We prove  Proposition 2.1 in the next subsection on the base of recurrent
relations for
$D^{(q)}$ that we derive now.  
Let us use identity for centered random variables 
$\E \{X^o Y^o\} = \E \{X Y^o\}$ and consider equality
$$
 \E \left\{ L^o_{a_1} L^o_{a_2} \cdots
L^o_{a_q}\right\} =\E \left\{ L_{a_1} [L^o_{a_2} \cdots L^o_{a_q}]^o\right\}
= {1\over N} \sum_{x,s=1}^N\E \left\{ H_{xs} (H^{a_1-1}_{sx} [L^o_{a_2} \cdots L^o_{a_q}]^o\right\}
\eqno (2.17)
$$
We apply to the last expression the integration by parts formula (5.1) 
and obtain equality
$$
D^{(q)}_{a_1,\dots,a_q} = 
{1\over 4} \sum_{j=0}^{a_1-2} 
\E \left\{ L_{a_1-2-j} L_{j} [ L^o_{a_2} \cdots L^o_{a_q}]^o \right\} 
$$
$$
+{1\over 4 N^2}\sum_{i=2}^q 
\E \left\{ L^o_{a_2} \cdots L^o_{a_{i-1}} 
\ a_i \ L_{a_i+a_1-2}L^o_{a_{i+1}} \dots L^o_{a_q}\right\},
\eqno (2.18)
$$
with the help of  formulas
(5.7) and  (5.8), respectively. The detailed derivation of (2.18) is presented in subsection 5.2.

Let us consider the first term from the right-hand side of (2.18).
We can rewrite it in terms of variables $D$ with the help of the following
identity
$$
\E \{L_1 L_2 Q^o\} = \E \{L_1\} \E \{L_2^o Q\} +
\E\{L_2\} \E \{L_1^o  Q\}+
\E \{L_1^o L_2^o Q\}-
\E \{L_1^o L_2^o \} \E \{Q\},
$$
where $Q = L_{a_2}^o \cdots L_{a_q}^o$. 
Regarding the last term of (2.18), we use (2.17) and
 obtain  relation
$$
D^{(q)}_{a_1,\dots,a_q} = {1\over 4}\sum_{j=0}^{a_1-2} M_j
D^{(q)}_{a_1-2-j,a_2,\dots,a_q}+
{1\over 4}\sum_{j=0}^{a_1-2} M_{a_1-2-j}
D^{(q)}_{j,a_2,\dots,a_q}
$$
$$
+{1\over 4} \sum_{j=0}^{a_1-2} D^{(q+1)}_{j, a_1-2-j,a_2,\dots,a_q}-
{1\over 4} \sum_{j=0}^{a_1-2} D^{(2)}_{j, a_1-2-j} D^{(q-1)}_{a_2,\dots,a_q}
$$
$$
+{1\over 4 N^2} \sum_{i=2}^q a_i M_{a_1 +a_i-2}
D^{(q-2)}_{a_2,\dots,a_{i-1}, a_{i+1},\dots,a_q} +
{1\over 4 N^2}\sum_{i=2}^q a_i D^{(q-1)}_{a_2,\dots,a_{i-1},
a_i+a_1-2, a_{i+1},\dots,a_q}.
\eqno (2.19)
$$
Taking into account that $M_{2k+1}^{(N)} = 0$, it is easy to deduce from (2.19) by 
induction on $k$   that 
$$
D_{a_1, \dots, a_q}^{(q)} = 0
\quad \hbox{whenever} \ \ 
a_1 + \dots +a_q = 2k+1.
$$
Let us introduce variables 
$$
\bar D^{(q)}_{2k} = \sum_{a_1+\dots+a_q=2k}
\left| D^{(q)}_{a_1,\dots, a_q} \right|.
$$
Using the positivity of $M_{2j}$,
we derive from (2.19) the second main relation 
$$
\bar D_{2k}^{(q)} \le {1\over 2} \sum_{j=0}^{k-1} \bar D^{(q)} _{2k-2-2j} M_{2j} +
{q-1\over 4N^2} \sum_{j=0}^{k-1} \bar D_{2k-2-2j}^{(q-2)}\cdot 
{(2j+2) (2j+1)\over 2} \cdot M_{2j}
$$
$$
+{1\over 4} \bar D^{(q+1)}_{2k-2} + 
{1\over 4} \sum_{j=0}^{k-1}\bar D^{(q-1)}_{2k-2-2j} \bar D^{(2)}_{2j} +
{2k(2k-1)\over 2} \cdot{ (q-1)\over 4N^2} \cdot  \bar
D_{2k-2}^{(q-1)}, 
\eqno (2.20)
$$
where $1\le k, 2\le q\le 2k$. 
When regarding two last terms of (2.19), we have used obvious equality
$$
\sum_{a_1+a_2=a'} a_2 \ F_{a_1+a_2-2} = 
\left(\sum_{a_2=1}^{a'-1} a_2\right)F_{a'-2} = 
{a'(a'-1)\over 2}
F_{a'-2}.
$$
Using this relation with $F$ replaced by $M$ and $a'=2j+2$, we obtain that
$$
\sum_{a_1+\dots+a_q=2k} a_i M_{a_1+a_i-2} \vert D^{(q)}_{a_1, \dots, a_{i-1},
a_{i+1}, \dots , a_{q}} \vert =  \sum_{j=0}^{k-1} \bar D_{2k-2-2j}^{(q-2)} 
{(2j+2) (2j+1)\over 2} \cdot M_{2j}.
$$
Also we can write that
$$
\sum_{a_1+\dots+a_q = 2k} 
a_i \vert D^{(q-1)}_{a_2,\dots,a_{i-1},
a_i+a_1-2, a_{i+1},\dots,a_q}\vert  
 $$
$$
=\sum_{b_2+\dots+b_{q}=2k-2} \vert D^{(q-1)}_{b_2, \dots, b_{q}} \vert 
\times \sum_{1\le a_1\le b_{i}+1} (b_i-a_1+2) \le  \sum_{a_1+\dots+a_{q-1}=2k-2} \vert D^{(q-1)}_{a_1, \dots, a_{q-1}}\vert \times {2k(2k-1)\over 2}
$$
and get the last term of (2.20).

The upper bounds of sums in (2.20) are written under agreement that 
$\bar D_{2k}^{(q)} = 0$ whenever $q> 2k$. Also we note that the form of inequalities
(2.20) is slightly  different when we consider particular values of $q$ and $k$.
Indeed,
some terms are missing when 
the left-hand side is  $\bar D^{(2)}_{2k}$,
$\bar D^{(3)}_{2k}$, $\bar D^{(2k)}_{2k}$,
$\bar D^{(2k-1)}_{2k}$,
and $\bar D^{(2k-2)}_{2k}$. However, the agreement that 
$\bar D_{2k}^{(q)} =0$ whenever $q>2k$
and that $\bar D^{(1)}_{2k} = 0$ and $ \bar D^{(0)}_{2k} = \delta_{k,0}$ make
(2.20) valid in these cases. 

Obviously, we have that
$$
M_{2k} \le {1\over 4} \sum_{j=0}^{k-1} M_{2k-2-2j}\, M_{2j} + {1\over 4} \bar D_{2k-2}^{(2)}.
\eqno (2.21)
$$

\subsection{Recurrent relations and estimates}

To estimate $M$ and $\bar D^{(q)}$, we introduce auxiliary numbers $B_k^{(N)}\ge 0$ and
$R^{(q;N)}_{2k}\ge 0$ determined by a system of two recurrent relations  induced by  (2.20) and (2.21).
This system is given by the following equalities (we omit superscripts $N$)
$$
B_k = {1\over 4} \left(B*B\right)_{k-1} + {1\over 4} R_{k-1}^{(2)}, 
\eqno (2.22)
$$
and 
$$
R_k^{(q)} = {1\over 2} \left( R^{(q)} * B\right)_{k-1} + {q-1\over 4N^2}
\left( R^{(q-2)}*B''\right)_{k-1}+{1\over 4} R^{(q+1)}_{k-1}
$$
$$
 + {1\over 4} \left( R^{(q-1)}*R^{(2)}\right)_{k-1} +
{k^2q\over 2N^2} R_{k-1}^{(q-1)}, 
\eqno (2.23)
$$ 
considered in the domain 
$$
\Delta = \{( k,q): \ k\ge 1, \, 2\le  q \le 2k\}
$$
with  denotation
$$
B''_k = {(2k+2)(2k+1)\over 2} B_k
$$
and the  convolutions  as follows
$$
\left( B*B\right)_{k-1} = \sum_{j=0}^{k-1} B_{k-1-j} B_j.
$$
The initial values for (2.22)-(2.23) coincide with those of $M$ and $D$: 
$$
B_0^{(N)} = 1,\  
\quad R_{1}^{(2;N)} = {1\over 4N^2}.
$$
Let us note that one can consider relations (2.22) and (2.23) for all integers  $k$ and $q$
with obvious agreement that outside of $\Delta$ the values of $R$ are zero
except the origin $R^{(0;N)}_0 = 1$.
The system  (2.22)-(2.23) plays a fundamental role in our method  
the proof of Proposition 2.1. 
This proof is composed of the following three statements. 
\vskip 0.5cm

\noindent {\bf Lemma 2.1}. 

\noindent {\it Given fixed $N$, the family of numbers $\{B_{k}, R^{(q)}_k, \ (k,q)\in \Delta\}$ exist; it is  uniquely determined  by the system of relations (2.22)-(2.23).}

\vskip 0.5cm

\noindent {\bf Lemma 2.2}. 

\noindent {\it Inequalities
$$
M_{2k}^{(N)} \le B_k^{(N)} \quad {\hbox{and }} \quad \bar D^{(q;N)}_{2k} \le R^{(q;N)}_{k}
\eqno (2.24)
$$
hold for all $N$ and  $(k,q)\in \Delta$.}

\vskip 0.5cm

\noindent {\bf Lemma 2.3}. 

\noindent{\it 
Under conditions of Proposition 2.1, the numbers $B_k$ and $R^{(q)}_{k}$
are estimated by the right-hand sides of 
inequalities (2.10), (2.14) and (2.15), respectively; that is 
$$
B_k^{(N)} \le \left[ f(\tau) + A N^{-2} \tau^2(1-\tau)^{-5/2}\right]_k \equiv [ \Phi_N(\tau)]_k
\eqno (2.25)
$$
and
$$
R^{(q;N)}_{k} \le 
\cases{C(3s)! N^{{-2s}}\left[ \tau (1-\tau)^{-2s}\right]_k, & if $q=2s$;\cr
C(3s+3)! N^{-2s-2} \left[\tau  (1- \tau)^{-(4s+5)/2}\right]_k, & if $q=2s+1$.\cr}
\eqno (2.26)
$$
}

\vskip 0.5cm
Lemma 2.3 represents the main technical  result concerning the system (2.22)-(2.23).
Lemmas 2.1 and 2.2 looks like a simple consequence of the recurrent procedure applied to
relations (2.22)-(2.23) and (2.20)-(2.21), respectively. However, the form of recurrent relations (2.22)-(2.23) is not usual because relations for $B$ involve the values of $R$ and vice-versa. The ordinary scheme of recurrence has to be modified. This modification is described in the next subsection.
Lemma 2.3 is also proved on the base of this modified scheme of recurrence.

\vskip 1cm
\subsubsection{The triangular scheme of recurrent estimates}

Let us show on the example of Lemma 3 that the ordinary scheme of recurrent estimates
can be applied to the system  (2.20)-(2.23).
Under the ordinary scheme we mean the following reasoning. Assume that the estimates
we need are valid for the terms entering the right-hand side
of the inequalities derived. Apply these estimates to all terms there and show that the sum
of the expressions obtained is smaller than that we assume for the terms of the 
left-hand side; check the estimates of the initial terms.
Then all estimates we need are true.
Let consider the plane of integers 
$(k,q)$ assume that estimates (2.26) are valid for all variables $R$
with $(k,q)$ 
lying inside of the triangle domain $\Delta(m), m\ge 3$ 
$$
\Delta(m) = \{ (k,q): \ 1\le k,\,  2 \le q\le 2k, \,
k+q \le m\}
$$
and that estimates (2.25) are valid for all 
variables $B_{l}$ with $1\le l \le m-2$.


\begin{figure}[htbp]
\centerline{\includegraphics[width=12cm]{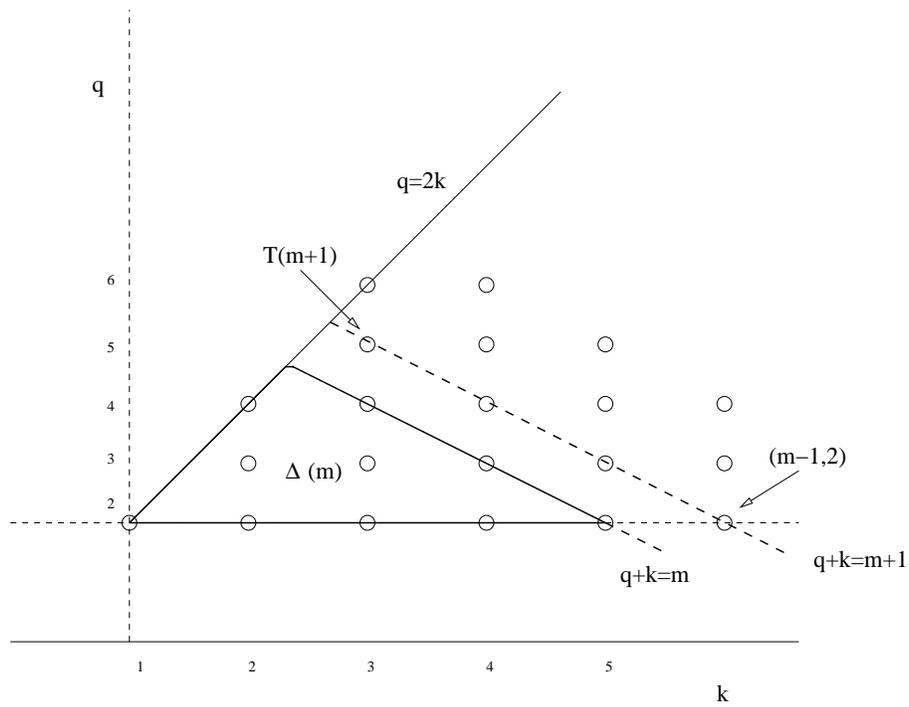}}
\caption{The triangle domain $\Delta(m)$ with $m=7$ and the long dotted line $k+q=m+1$}
\end{figure}


Then we proceed to complete the next line $k+q = m+1$ step by step 
starting from the top point $T(m+1)$ of  $\Delta(m+1)$ and
ending at the bottom point 
$(m-1,2)$ of this side line.  
This means that on each step, we assume estimates (2.25) and (2.26) valid
for all terms entering the right-hand sides of relations (2.23) and show that
the same estimate is valid for the term standing at the left hand side
of (2.23). 

Once the bottom  point $(m-1,2)$ achieved, we turn to relation (2.22)
and prove that estimate (2.25) is valid for $B_m$. Again, 
this is done by assuming that all terms entering the right-hand side
of (2.22) verify estimates (2.25) and (2.26) with $q=2$, and showing
that the expression obtained is bounded by the right-hand side of (2.25).
This completes the triangular scheme of recurrent estimates.

It is easy to see that the reasoning 
described above proves, with obvious changes,  Lemmas 2.1 and 2.2.

\vskip 1cm

\subsubsection{Estimates for $B$}

Assuming that the terms standing 
in the right-hand side of (2.22) are 
estimates (2.25) and (2.26) with $s=1$,
we can write  inequality
$$
{1\over 4} (B*B)_{k-1} + {1\over 4} R_{k-1}^{(2)}  
$$
$$
\le \left[{\tau f^2(\tau )\over 4} + 
{A\over N^2}{\tau^3 f(\tau)\over
2(1-\tau)^{5/2}} 
+ {A^2\over 4N^4} {\tau^5 \over (1-\tau)^5}\right]_k + {3C\over 2 N^2}
\left[{\tau^2\over (1-\tau)^2}\right]_k.
\eqno (2.27)
$$

Taking into account relations 
(2.11),
we  transform the first bracket of (2.27) into expression
$$
\left[ f(\tau) + {A\over N^2} {\tau^2\over (1-\tau)^{5/2} } \right]_k - 
{A\over N^2} \left[{\tau^2 \over (1-\tau)^2} \right]_k + {A^2\over 4N^4} 
\left[{\tau^5\over (1-\tau)^5} \right]_k.
$$
Here, the first term reproduces expression $[\Phi_N(\tau)]_k$ ; the
second term is negative and this allows us to show that the estimate wanted is true.
Then we see that estimate
$B_k \le  [\Phi_N(\tau)]_k$
is true whenever inequality
$$
A \left[ {\tau^2\over (1-\tau)^2}\right]_k \ge
 {3C\over 2} \left[ {\tau^2\over (1-\tau)^2}\right]_k +
{A^2\over 4N^2} \left[ {\tau^5 \over (1-\tau)^5}\right]_k
\eqno (2.28)
$$
holds. This is equivalent to the condition
$$
A \ge {3C\over 2}  + {A^2\over 4N^2} {(k-4)(k-3)(k-2)\over 4!}.
$$
Remembering that  $k^3\le \chi N^2$, we see that the estimate (2.25) of $B_k$
is true provided
$$
A \ge {3C\over 2} + {A^2\chi\over 96}.
\eqno (2.29)
$$

\vskip 1cm

\subsubsection{Estimates for $R^{(2s)}$}
Let us rewrite (2.8) with $q= 2s, s\ge 2, \ k\ge 1$ in the form
$$
R_k^{(2s)} = {1\over 2} \left( R^{(2s)} * B\right)_{k-1} + 
{2s-1\over 4N^2} \left( R^{(2s-2)} * B''\right)_{k-1}
+
X+Y+Z,
\eqno (2.30)
$$
where we denoted
$$
X = {1\over 4} R^{(2s+1)}_{k-1}, 
\quad Y = {1\over 4} \left(
R^{(2s-1)}*R^{(2)}\right)_{k-1},\quad 
 Z ={k^2 s\over N^2} R_{k-1}^{(2s-1)}.
\eqno (2.31)
$$ 

The first term in the right-hand side of
(2.30) admits the following estimate
$$
{1\over 2} \left( R^{(2s)}*B\right)_{k-1} \le
 {C(3s)!\over N^{2s}} \left[ {\tau^2 f(\tau)\over 2(1-\tau)^{2s} }
 + {A\over 2 N^2} {\tau^4 \over (1-\tau )^{2s+ 5/2}}  \right]_k.
$$
Using (2.11), we transform the last expression to the form 
$$
{C(3s)!\over N^{2s}} \left[{\tau\over (1-\tau)^{2s}} - 
 {\tau\over (1-\tau)^{2s-1/2}} +
{A \over 2 N^{2}} {\tau^4\over (1-\tau)^{2s+5/2}}\right]_k.
\eqno (2.32)
$$ 
The first term reproduces the expression
we need to estimate $R^{(2s)}_k$.

Let us consider the second terms of the right-hand side of (2.30).
Assuming (2.25) and using identities of subsection 5.1, it is not hard  to show that
$$
B_k'' \le \left[ {1\over (1-\tau)^{3/2}}  +
{18 A\over  N^2} {1\over  (1-\tau)^{9/2}}\right]_k.
\eqno (2.33)
$$
Indeed, it follows from (5.11) that
$$
{(2k+2)(2k+1)\over 2} [f(\tau)]_k = {(2k+2)(2k+1)\over 2} m_k = \left[ {1\over (1-\tau)^{3/2}}  \right]_k.
$$
Next, identity (5.12) implies relation
$$
{(2k+2)(2k+1)\over 2} \left[{\tau^2\over  (1-\tau)^{5/2}}\right]_k = 
{(2k+2)(2k+1)\over 2}\cdot {2k (2k-1)(2k+1)\over 3!} m_k.
$$
Now, regarding (5.9) with $r=4$, it is easy to see that
$$
\left[ {1\over (1-\tau)^{9/2}} \right]_k = {(2k+1)(2k+2)(2k+3)(2k+5)(2k+7)\over 5\cdot 6\cdot 7} m_k.
$$
Then (2.33) follows.

Returning to the right-hand side of (2.30), we can write with the help of (2.33) inequality
$$
 \left( R^{(2s-2)}* B''\right)_{k-1} \le 
{C(3s-3)! \over 4N^{2s-2}} 
 \left[ {\tau\over (1-\tau)^{2s-1/2}} 
+
{18A\over N^2} 
 {\tau\over (1-\tau)^{2s+5/2}}\right]_k.
\eqno (2.34)
$$
Here and below we use relation $[\tau^j g(\tau)]_k\le [g(\tau)]_k$ valid
for the generating  functions under consideration.
Let us stress that (2.34) remains valid in the case of $s=1$
with $C$ replaced by $1$.

Let us turn to (2.31).
We estimate the sum of $X$ and $Y$ by 
$$
X+Y \le {C(1+C)(3s+3)!\over 4N^{2s+2}} \left[
{\tau \over (1-\tau)^{2s+5/2}}\right]_k\ .
\eqno (2.35)
$$
For the last term of (2.31) we can write inequality
$$
Z \le
{C k^2 (3s+1)! \over N^{2s+2}} \left[ {\tau \over (1-\tau)^{2s+1/2}}\right]_k
\ .
\eqno (2.36)
$$

Comparing the second term of (2.32) with the sum of the last term of (2.32)
and the right-hand sides of (2.34), (2.35), and (2.37), 
 we
arrive at the following inequality to hold
$$
C \ge {(2s-1)(3s-3)!\over (3s)!} \cdot {\delta_{s,1} + C(1-\delta_{s,1}) \over 4} +
{k^2 (3s+1)\over 3N^2}\cdot {\left[\tau (1-\tau)^{-2s- 1/2}\right]_k\over 
\left[ \tau
(1-\tau)^{-2s+1/2}\right]_k}\ 
$$
$$
+ C{(1+C)(3s+3)! + 
18A(3s-2)! + 2A (3s)!\over 4N^2(3s)!}\cdot 
{\left[ \tau (1-\tau)^{-2s- 5/2}\right]_k\over 
\left[\tau 
(1-\tau)^{-2s+1/2}\right]_k}\ .
\eqno (2.37)
$$
Using identity (5.10), we see that
$$
{\left[ \tau (1-\tau)^{-2s- 1/2}\right]_k\over 
\left[\tau 
(1-\tau)^{-2s+1/2}\right]_k} = {2k+4s-2\over 4s-1} \le {4k_0\over 4s-1}.
$$
Similarly
$$
{\left[ \tau (1-\tau)^{-2s- 5/2}\right]_k\over 
\left[\tau 
(1-\tau)^{-2s+1/2}\right]_k}\  \le {(4k_0)^3\over (4s-1)(4s+1)(4s+3)}.
$$
Inserting these inequalities  into (2.37),maximizing 
the expressions obtained with respect to $s$, and using (2.13),  we get the following 
sufficient condition
$$
C \ge {\delta_{s,1} + C(1-\delta_{s,1})\over 24} 
+
2\chi \, \left( 1 + 10 C (1+C) + 2AC\right).
\eqno (2.38)
$$

\subsubsection{Estimates for $R^{(2s+1)}$}

Let us turn to the case $q= 2s+1$ and rewrite (2.8) in the form
$$
R_k^{(2s+1)} = {1\over 2} \left( R^{(2s+1)} * B\right)_{k-1} + {s\over 2N^2}
\left( R^{(2s-1)}*B''\right)_{k-1}+ 
X_1 + Y_1 + Z_1,
\eqno (2.39)
$$
where 
$$
X_1 = {1\over 4} R^{(2s+2)}_{k-1},
\quad  Y_1 =  {1\over 4} \left(
R^{(2s-1)}*R^{(2)}\right)_{k-1},\ 
Z_1 = {k^2 s\over N^2} R_{k-1}^{(2s)}.
\eqno (2.40)
$$ 
Regarding the first term of
(2.39), we can write inequality
$$
{1\over 2} \left( R^{(2s+1)}*B\right)_k \le 
{C(3s+3)!\over  N^{2s+2}}
\left[ {\tau^2 f(\tau) \over 2 (1-\tau)^{2s+5/2} } +{A \tau \over 2N^2 (1-\tau)^{2s+5}}
\right]_k 
$$
$$
={C(3s+3)!\over N^{2s+2} }\left[{\tau\over (1-\tau)^{2s+5/2}}  -
{\tau \over (1-\tau)^{2s+2}} +
{A\tau \over 2N^2 (1-\tau)^{2s+5}} \right]_k\ . 
\eqno (2.41)
$$
The first term of the right-hand side of (2.41)
reproduces the expression needed to estimate $R^{(2s+1)}_k$.

Let us consider the second term of (2.39). It  is estimated as follows:
$$
{s\over 2N^2}
\left( R^{(2s-1)}*B''\right)_{k-1} \le 
{C s (3s)! \over 2 N^{2s+2} } \left[{\tau\over (1-\tau)^{2s+2}} \right]_k +
{9 A C s (3s)! \over N^{2s+4} } 
\left[{\tau \over (1-\tau)^{2s+5}} \right]_k \ .
$$
 
Regarding two first terms of (2.40), we can write that 
$$
X_1 +  Y_1 \le  
{C(3s+3)! + 6C^2(3s)!  \over 4N^{2s+2} } 
\left[ \tau\over (1-\tau)^{2s+2}\right]_k,
$$
and 
$$
 Z_1 \le  {C k^2 s(3s)!\over  N^{2s+2} } 
 \left[{\tau \over (1-\tau)^{2s}} \right]_k.
$$
Comparing the negative term of (2.41) with the sum of the last term of (2.41)
and the estimates for the terms of (2.40), we obtain inequality
$$
C\left( {3\over 4} - {s(3s)!\over 2(3s+3)!}\right) \ge
{3C^2(3s)!\over  2(3s+3)!} + 
{ k^2 s(3s)!\over (3s+3)!} \cdot 
{[\tau (1-\tau)^{-2s}]_k\over [\tau (1-\tau)^{-2s-2}]_k}\ 
$$
$$
+{AC\over 2N^2}\cdot  \left(1 + {18s(3s)!\over (3s+3)!} \right)
\cdot { [\tau (1-\tau)^{-2s-5}]_k\over  [\tau (1-\tau)^{-2s-2}]_k}.
\eqno (2.42)
$$ 
Equality (5.13) implies that
$$
{[\tau (1-\tau)^{-2s}]_k\over [\tau (1-\tau)^{-2s-2}]_k}\ = {2s(2s+1)\over (k-1+2s)(k+2s)}
$$
and that
$$
{ [\tau (1-\tau)^{-2s-5}]_k\over  [\tau (1-\tau)^{-2s-2}]_k} 
 \le 
{ 8k_0^3\over (2s+2)(2s+3)(2s+4)}.
$$
Inserting these two relations into (2.42) and maximizing expressions with respect to $s$,
we obtain, after elementary transformations, the following sufficient
condition
$$
C \le {4!\over 1 + 4A\chi}.
\eqno (2.43)
$$

\vskip 0.5cm

\subsection{Proof of Theorem 2.1}

Let us repeat that inequalities (2.29), (2.38), and (2.43) represent sufficient conditions
for recurrent estimates (2.25) and (2.26) to be true.  
Let $A>1/16$. Then for any constant $C<4!$ verifying condition
$$
{1\over 24} < C < {2A\over 3},
$$
there exists such $\chi >0$ that (2.38) is true. Indeed, it is sufficient to take 
$\chi\le \chi'$, where $\chi'$ is such that 
$$
2\chi' K < \min\{ C - {1\over 24}, {23\over 24}C\}, 
$$
with $ K = 1+ 10 C(1+C) +2AC$.
Also there exists $\chi''$ such that (cf. (2.29))
$$
A\ge 3C/2 + A^2 \chi''.
$$
The choice of  $\chi\le \min\{\chi',\chi''\}$ makes (2.29) and (2.38) true. Condition (2.43)
is obviously verified.
Thus, 
conditions (2.13), (2.14),
and $A>1/16$  of Proposition 2.1 are sufficient for (2.29), (2.38), and (2.43) to hold.
This completes the proof of Lemma 2.3.

Lemma 2.2 together with Lemma 2.3 implies estimates (2.10) and (2.15).
Then Proposition 2.1 follows. The statement of Theorem 1.1 is a simple 
consequence of the estimate (2.10) and Proposition 2.1.

\vskip 1cm

We complete this subsection with the discussion of the form of estimates (2.26)
and constants $A$ and $C$.
 First let us note that the upper bound $4!$ for $C$ imposed by (2.14)  represents a
technical restriction; it can be avoided, for example, by modifying estimates (2.26) for
$R^{(2s)}$ and
$R^{(2s+1)}$, where $C$ is replaced by $C^s$ and $C^{s+1}$, respectively. 
However, in this case the lower bounds
$1/16$ for $A$ and $1/24$ for $C$ are  to be replaced by $1/6$ and $1/9$,
respectively.

The closer $A$ and $C$ to optimal values $1/16$ and $1/24$ are, the smaller $\chi$ is to be
chosen. 
The inverse is also correct. Namely,  in the next subsection we prove that
estimates (2.9) and (2.10)
become asymptotically exact
 in the limit $\chi \to 0$.
In this case factorials $(3s)!$ and $(3s+3)!$ in the right-hand sides of (2.26)
can be replaced by other expressions $g(s)$ and $h(s)$ that provide more precise estimates
for
$R^{(q)}$. Indeed, repeating the computations
of subsections 2.3.3 and 2.3.4, one can see that in the limit $\chi\to 0$
function $g(s)$ can be chosen close to $(2s-1)!!/4^s$. 
This make an evidence for the central limit theorem to hold
for the centered random variables
$$
N L_a^o = \T\, H^a - \E \{\T\, H^a\}.
$$ 
This observation explains also the fact that the odd "moments" of the variable $L^o_a$
decrease faster than the even ones as $N\to\infty$. That is why the estimates for
$R^{(2s)}$ have the form different from those of $R^{(2s+1)}$ and are proved separately.

For finite values of $\chi$, the use of some  expression proportional to $(3s)!$
is unavoidable.


\subsection{ Proof of Theorem 2.2}
We present the proof of Theorem 2.2 for the case when $k$ is fixed and $N\to\infty$.
Regarding  relation (2.19) with $q=2$, we obtain relation
$$
D^{(2)}_{2k} = {1\over 2} \left( D^{(2)}*M\right)_{2k-2} + {1\over
4N^2}\cdot {2k(2k-1)\over 2} M_{2k-2} + {1\over 4} D^{(3)}_{2k-2}.
\eqno (2.44)
$$
Proposition 2.1 implies  that $D^{(3)}_{2k} = O(N^{-4})$ and that
$M_{2k}^{(N)} - m_k = O(N^{-2})$. Then we easily arrive at the conclusion that  
$$
D^{(2)}_{2k} = {r_k\over N^2} + O\left({1\over N^4}\right),
\eqno (2.45)
$$
where  $r_k$ are determined by relations $r_0=0$ and 
$$
r_k = {1\over 2} \left( r*m\right)_{k-1} + 
{1\over 4}\cdot {2k(2k-1)\over 2} m_{k-1}, \quad k\ge 1.
\eqno (2.46)
$$
Passing to the generating functions and using relations (2.11) and 
(5.11), we obtain equality
$$
r_k = {1\over 4} \left[{\tau\over (1-\tau)^2}\right]_k = {k\over 4}.
$$

Returning to relation (2.8), we conclude that 
$$
M_{2k}^{(N)} = m_k + {1\over N^2 } m_k^{(2)} + 
O\left({1\over N^4}\right).
$$
Indeed, the difference between $M_{2k}^{(N)}$ and $m_k$ is of the order $N^{-2}$
and the next correction is of the order $N^{-4}$. Regarding $m_k^{(2)}$ and using (2.45),
we obtain equality
$$
m_k^{(2)} = {1\over 2} \left[ m^{(2)} * m\right]_{k-1} + {1\over 16}
\left[ {\tau^2\over (1-\tau)^2}\right]_k, \quad k\ge 1,
\eqno (2.47)
$$
and $m_1^{(2)} = 0$. Solving (2.47) with the help of (2.11), we get
expression
$$
m_k^{(2)} = {1\over 16} \left[ {\tau^2 \over (1-\tau)^{5/2}} \right]_k.
\eqno (2.48)
$$
It is easy to see that  (2.48) implies relation 
$$
{1\over 16} \left[ {\tau^2 \over (1-\tau)^{5/2}} \right]_k = 
{1\over 16} {(2k-3)(2k-2)(2k-1)\over 3!} m_{k-2}
$$
and hence  (2.6b). Theorem 2.2 is proved.

\subsection{More about asymptotic expansions}

The system (2.22)-(2.23) of recurrent relations is the main technical tool in the proof of 
the Proposition 2.1, where the estimates for $B$ and $R$ are given. 
However, the crucial question  is to find the correct form of these estimates.
The first terms of the asymptotic expansions 
described in previous subsection give a solution of this problem.
Indeed, repeating the proof of Theorem 2.2, we see that formulas (2.46) and (2.48)
indicate the form of the estimates to be proved.
Then the proof of Proposition 2.1 is reduced to elementary computations, where the most important part is related with the correct choice of the factorial terms in 
inequalities (2.15).

The next observation is  that relation (2.23) resembles inequality (2.20) 
obtained from (2.19) by considering the absolute values of variables 
$D^{(q)} _{a_1, \dots,  a_q}$ and replacing in the right-hand side of (2.19)
the sign "$-$" by the sign "+". So, relation (2.23) determine the estimating terms $R^{(q)}$
with certain error. However, it is not difficult to deduce from estimates (2.25) and (2.26) that if $q=2s$, then  this error is of the order smaller than the order  of  $R^{(2s)}$. This means that relations (2.23) determine
correctly the first terms of the $1/N$-expansions of all $R^{(2s)}, s\ge 1$ and not only of 
$R^{(2)}$ as mentioned by Theorem 2.2. The same is true for the $1/N$ expansions of 
$D^{(2s)}_{2k}$. It is easy to show by using (2.23) and results of Proposition 2.1 that these corrections are
given by formulas
$$
D^{(2s)}_{2k} = r^{(2s)}_k + o(k^{2s-1}/N^{2s}), 
$$
where $r^{(2s)}_k$ are such that the corresponding generating function 
$\tilde r^{(2s)}(\tau) = \sum_{k\ge 0} r^{(2s)}_k \tau ^k$ verifies equation
$$
\tilde r^{(2s)}(\tau ) = {\tau f(\tau)\over 2} \tilde r^{(2s)}(\tau ) +
(2s-1) \tilde r^{(2s-2)}(\tau) {d^2 \over 2 N^2 d\tau^2} ( \tau f(\tau)).
\eqno (2.49)
$$
Using equalities (2.11) and resolving (2.49), we obtain expression 
$$
\quad r^{(2s)}_k =  {(2s-1)!!\over (4N^2)^s }
\left[ {\tau^s\over (1-\tau)^{2s} }\right]_k.
$$

The left-hand side of relation (2.23) for $R^{(q)}_k$ involves 
variables $R^{(q)}_j$, $R^{(q-1)}_j$, and $R^{(q+1)}$. This can lead one to the idea to use the generating functions of two variables $G(\tau,\mu)$  to describe the family of numbers $R$.
In this connection, the following comment on the structure of the variables  $D^{(q)}$ could be useful. Introducing
a generating function
$
F(\tau) = \sum_{j\ge 0} \tau^j L_j
$,
we see  that
$$
\sum_{k\ge 1} D^{(q)}_{2k} \tau ^{2k} = \E \{ [F^o(\tau)]^q\},
$$
where $F^o(\tau ) = F (\tau) - \E F(\tau)$.
Then the mentioned above function can have the form
$$
G_D(\tau,\mu) = \sum_{k\ge 1, q\ge 2} D^{(q)}_{2k} \tau^{2k} {\mu^q\over q!} = 
\E\left\{ e^{\mu F^o(\tau)}\right\} -1.
$$
In particular, regarding such a  generating function of $r^{(2s)}_{k}$, one arrives at the expression
$$
G_r(\tau,\mu) =  \sum_{k\ge 1, s\ge 1} r^{(2s)}_{k} \tau^{2k} {\mu^{2s}\over (2s)!} = 
\exp\left\{{\mu^2\over 4N^2}  {\tau\over  (1-\tau)^2}\right\}.
$$
This expression show that the central limit theorem can be proved for the random variable $NF^o(\tau)$ in the asymptotic regime $ k^3/N^2\ll 1$ mentioned in Theorem 2.2. This asymptotic regime can be compared with the mesoscopic regime for the resolvent of $H_N$ 
and the central limit theorem valid there \cite{BK1}.




\section{Orthogonal and anti-symmetric ensembles}

In this section we return to   Hermitian random matrix ensembles
$H^{(\eta)}$ with $\eta = 1$ and $\eta = -1$ introduced in section 2. 
Let us consider the moments of $H^{(1)}$. Using the method developed in section 2,
we prove the following statements.

\vskip .5cm
\noindent  {\bf Theorem 3.1 (GOE).} 

\noindent {\it  Given $A>1/2 $, there exists
$\chi$  such  that 
$$
 M_{2k}^{(N)} \le m_k + A{1\over N}
\eqno (3.1)
$$
for all $k, N $ such that $k\le k_0$ and (2.13) hold. 
 If $k$ is fixed and
$N\to\infty$,
then
$$
 M_{2k}^{(N)} = m_k +  { 1 - (k+1)m_k\over 2 N}  + o(N^{-1})
\eqno (3.2)
$$
and
$$
D^{(2;N)}_{2k} = \sum_{a+b = 2k} 
 \E \left\{L_a^o L_b^o
\right\} = {k\over 2N^2} + O(N^{-3}).
\eqno (3.3)
$$
}

\vskip 0.5cm
The proof of Theorem 3.1 is obtained by 
using the method described in section 2. Briefly saying, we derive
recurrent inequalities for  $M_{2k}^{(N)}$ and $D^{(q)}_{2k}$, then introduce related
auxiliary numbers $B$ and $R$ determined by a system of recurrent 
relations. Using the triangular scheme of recurrent estimates to prove the estimates we
need.
Corresponding computations are somehow different from those of section 2. We describe this
difference below (see subsection 3.1).

Let us turn to  the ensemble $H^{(-1)}$. Regarding the recurrent relations
for the moments of these matrices, we will see that
for $M_{2k}^{(\eta = -1)}$ are bounded by $M_{2k}^{(\eta=1)}$. 
Slightly modifying the computations performed in the proof of Theorem 3.1,
one can prove the following result.

\vskip 0.5cm
\noindent{\bf Theorem 3.2 (Gaussian anti-symmetric Hermitian matrices). }

\noindent{\it Given $A > 1/2$, there exists $\chi>0$ 
such that the moments of Gaussian  skew-symmetric Hermitian ensemble $H^{(-1)}_N$
admit the estimate
$$
M_{2k}^{(N)} \le m_k + A{1\over N}
\eqno (3.4)
$$
for all values of $k,N$ such that (2.13) holds. Also 
$$
\vert D^{(2)}_{2k} \vert  = O\left( {1\over N^2}\right)
\quad \hbox{and} \quad 
\vert D^{(3)}_{2k}\vert = O\left({1\over N^3}\right).
\eqno (3.5)
$$
Given $k$ fixed, the following asymptotic expansions
are true for the moments of $H^{(-1)}$
$$
M^{(N)}_{2k} = m_k + {\delta_{k,0} - (k+1)m_k\over 2N} + o(N^{-1}),
\eqno (3.6)
$$
and for the covariance terms
$$
D^{(2;N)}_{2k} = \sum_{a_1+a_2=2k} \E \left\{L^o_{a_1} L^o_{a_2}\right\} = {k+1\over 4N^2} +
O(N^{-3}).
\eqno (3.7)
$$
}

\subsection{Proof of Theorem 3.1}

Using the integration by parts formula (5.7) with $\eta =1$  
and repeating computations of the previous
section,  we obtain recurrent relation for $M_{2k} = \E L_{2k}$;
$$
 M_{2k} = {1\over 4} \sum_{j=0}^{k-1}  M_{2k-2-2j}  M_{2j} +
 {2k-1\over 4N}  M_{2k-2} + {1\over 4} \sum_{a_1 + a_2 = 2k-2} 
\E \left\{  L_{a_1}^o  L_{a_2}^o\right\}.
\eqno (3.8)
$$
Regarding the variables
$$
D^{(q)}_{2k} = \sum_{a_1, \dots, a_q}^{2k} D^{(q)}_{a_1,\dots,a_q} = 
\sum_{a_1, \dots, a_q}^{2k} \E \left\{ L^o_{a_1} L^o_{a_2} \cdots
L^o_{a_q}\right\}
$$
and using  formulas (5.6) and (5.8) with $\eta = 1$,
we obtain relation
$$
D^{(q)}_{a_1,\dots,a_q} = {1\over 4}\sum_{j=0}^{a_1-2} M_j
D^{(q)}_{a_1-2-j,a_2,\dots,a_q}+
{1\over 4}\sum_{j=0}^{a_1-2} M_{a_1-2-j}
D^{(q)}_{j,a_2,\dots,a_q}
$$
$$
+{1\over 4} \sum_{j=0}^{a_1-2} D^{(q+1)}_{j, a_1-2-j,a_2,\dots,a_q}-
{1\over 4} \sum_{j=0}^{a_1-2} D^{(2)}_{j, a_1-2-j} D^{(q-1)}_{a_2,\dots,a_q}+
{1\over 4N} (a_1-1) \E \left\{ L_{a_1 - 2}^o L_{a_2}^o \cdots
L_{a_q}^o\right\}
$$
$$
+{1\over 2 N^2} \sum_{i=2}^q a_i M_{a_1 +a_i-2}
D^{(q-2)}_{a_2,\dots,a_{i-1}, a_{i+1},\dots,a_q} +
{1\over 2 N^2}\sum_{i=2}^q a_i D^{(q-1)}_{a_2,\dots,a_{i-1},
a_i+a_1-2, a_{i+1},\dots,a_q}.
\eqno (3.9)
$$
Introducing variables 
$$
\bar D^{(q)}_{2k} = \sum_{a_1,\dots, a_q}^{2k} 
\left| \E\left\{ L_{a_1}^o\cdots
L_{a_q}^o\right\}\right|,
$$
we derive from (3.9)  inequality
$$
\bar D_{2k}^{(q)} \le 
{1\over 2}  \sum_{j=0}^{k-1}\bar D^{(q)}_{2k-2-2j}M_{2j} +
{1\over 4} \bar D^{(q+1)}_{2k-2} + 
{1\over 4} \sum_{j=0}^{k-1} \bar D^{(q-1)}_{2k-2-2j} \bar D^{(2)}_{2j} +
{k\over 2N} \bar D^{(q)}_{2k-2}
$$
$$
 +
{q-1\over 2N^2} 
\sum_{j=0}^{k-1}  \bar D^{(q-2)}_{2k-2-2j}{(2j+2)(2j+1)\over 2} M_{2j}+
{ (q-1)k^2\over N^2} \cdot  \bar
D_{2k-2}^{(q-1)}.
\eqno (3.10)
$$
We have used here the same transformations as it was used when passing from equality (2.19) to inequality (2.20).

Now we proceed as in Section 2 and introduce the auxiliary numbers $ B$ and $ R$ that verify relations
$$
 B_k = {1\over 4} \left(  B*  B\right)_{k-1} + {k\over 2N}  B_{k-1} 
+ {1\over 4}  R^{(2)}_{k-1}, \quad k\ge 1,
\eqno (3.11)
$$
and
$$
 R^{(q)}_{k} = {1\over 2} \left(  B*  R^{(q)}\right)_{k-1} +
 {q-1\over 2N^2}
\left( B''* R^{(q-2)}\right)_{k-1-j}
$$
$$
+{ k\over 2N}  R_{k-1}^{(q)} + {1\over 4}  R^{(q+1)}_{k-1} 
+{1\over 4} \left(  R^{(2)}*  R^{(q-1)} \right)_{k-1}
+
 {q k^2\over N^2} R_{k-1}^{(q-1)}.
\eqno (3.12)
$$
The initial conditions are: 
$B_0 = 1$, $R_1^{(2)} = 1/(2N^2)$. 
The triangular scheme of recurrent estimates implies inequalities
$$
M_{2k}^{(N)} \le B_k^{(N)}, \quad {\hbox{and}}
\quad |D_{2k}^{(q)}|\le \bar D^{(q)}_{2k} \le R_k^{(q)}.
\eqno (3.13)
$$
The main technical result for GOE is given by the following proposition.
\vskip 0.5cm

\noindent {\bf Proposition 3.1}

\noindent {\it Let us consider $B$ and $R$ for the case of GOE ($\eta = 1$).
Given $A>1/2$ and $1/4<C< 2\cdot 6!$, there exists $\chi$ such that 
the following estimates 
$$
 B^{(N)}_k \le m_k + { A\over N},  \quad k\ge 2,
$$
or equivalently
$$
B^{(q)}_k \le \left[ f(\tau)\right]_k + { A \over N}
\left[ {\tau\over 1-\tau} \right]_k, \quad k\ge 2,
\eqno (3.14)
$$
and
$$
 R^{(2s)}_{k} \le { C(3s)! \over N^{2s} } \left[ {\tau\over
(1-\tau)^{2s}}
\right]_{k},
\eqno (3.15a)
$$
and
$$
R^{(2s+1)}_{k} \le {  C(3s+3)! \over N^{2s+2} } 
\left[ {\tau\over (1-\tau)^{2s+ 5/2}Ê} \right]_{k},
\eqno (3.15b)
$$
hold for all values of $k, q$ and $N $ such that $k\le k_0$ and (2.13) and (2.16) hold. 
}

\vskip 1cm The proof of this proposition resembles very much that of the Proposition 2.1.
However, there is a difference in the formulas that leads to somewhat different condition
on $A$.
To show  this, let us consider the estimate for $B_k$.
Substituting (3.14) and (3.15)
into the right-hand side of (3.11) and using (2.11),
we arrive at the following inequality (cf. (2.28))
$$
{ A \over N} \left[ {\tau \over \sqrt{1-\tau} } \right]_k \ge 
{k\over 2N} m_{k-1} + { A k \over 2N^2} \left[ {\tau\over 1- \tau } \right]_k 
+ {  A^2 +  6C\over 4N^2} \left[ {\tau^2\over (1-\tau)^2 } \right]_k
$$
that is sufficient for the estimate (3.14) to be true.
Taking into account that 
$$
\left[ {\tau\over \sqrt{1-\tau}} \right]_k = k m_{k-1},
\eqno (3.16)
$$
we obtain inequality
$$
A\ge {1\over 2}  + 
{2A + A^2 + 6C\over 4 N m_{k-1}}.
$$
It is easy to show that 
 $m_{k-1} \sqrt{k} \ge (2k)^{-1}$.
Then the last inequality is reduced to the condition 
$$
 A \ge {1\over 2}
+ (A +  A^2 +   3C) \sqrt{\chi}.
\eqno (3.17)
$$
The estimates for $R^{(q)}$ also include the values $\sqrt \chi$ and $\chi$.
We do not present these computations.

Let us prove the second part of Theorem 3.1.
Regarding relation (3.8) and 
taking into
account estimate (3.15a) with $q=2$, we conclude that 
$$
 M_{2k}^{(N)} = m_k + {1\over N}  m_k^{(1)} + o(N^{-1}), \quad 
{\hbox{as \ }} N\to\infty.
$$
It is easy to see that the numbers $ m_k^{(1)}$ are determined by relations
$$
 m_{k}^{(1)} = {1\over 2} \left(  m^{(1)} * m\right)_{k-1} + {2k-1\over 4}
m_{k-1}
\eqno (3.19)
$$
and $m^{(1)}_0 = 0$.
Passing to the generating functions, we deduce from (3.19)
equality
$$
m_k^{(1)} = {1\over 2} \left[ {\tau \over 1-\tau}\right]_k -
{1\over 2} \left[ {1 - \sqrt{1-\tau} \over \sqrt{1-\tau}} \right]_k
= {1\over 2} - {(k+1)m_k\over 2}.
$$
Relation (3.2) is proved.

Let us consider the covariance term $D^{(2)}$. 
It follows from the results of Proposition 3.1 that 
$$
D^{(2)}_{2k} = {r_k\over N^2} + o(N^{-2}).
$$
Then we deduce from (3.12) with $q=2$ that $r_k$ is determined by
the following recurrent relations
$$
r_k = {1\over 2} \left(r*m\right)_{k-1} + {1\over 2} {2k(2k-1)\over 2}
m_{k-1}.
$$
Solving this equation, we get
$$
r_k = \left[{\tau \over 2 (1-\tau)^2}\right]_k.
$$
This completes the proof of Theorem 3.1.

\subsection{Proof of Theorem 3.2}

In present section we consider the ensemble 
$H^{(\eta)}$ with $\eta = -1$. 
In this case the elements of $H$ (2.3) are given by 
imaginary numbers
$$
(H)_{xy} = {\i\over \sqrt N} W_{xy}, \quad x<y
$$
and the skew-symmetric condition holds:
$$
(H)_{xy} = - (H)_{yx} = (-H)_{yx}
$$
Regarding the last term of the formula (5.7) and using previous identity,
we can write that 
$$
{\eta\over 4N}\sum_{j=1}^{2k-1} \sum_{x,y=1}^N\E \left\{ (H^{j-1})_{yx}
(H^{2k-1-j})_{yx}
\right\} =
-{1\over 4} \sum_{j=1}^{2k-1} (-1)^{j-1} \E \left\{ {1\over N} \T H^{2k-2}
\right\}.
$$
Then 
we derive from (5.7) equality
$$
\E L_{2k} = {1\over 4} \sum_{j=0}^{2k-2} \E \left\{ L_{j} L_{2k-2-j}\right\}
-{1\over 4N} \E L_{2k-2}
$$
that gives recurrent relation for the moments of matrices $H^{(-1)}_N$
$$
M_{2k} = {1\over 4} \sum_{j=0}^{k-1} M_{2j} M_{2k-2-2j} -{1\over 4N} M_{2k-2} +
{1\over 4}D^{(2)}_{2k-2},
\eqno (3.20)
$$
where the term $D^{(2)}$ is determined as usual.
Regarding the general case of \mbox{$D^{(q)}, q\ge 2$,} 
and using  (5.8), we obtain the following relation
$$
D^{(q)}_{a_1,\dots,a_q} = 
{1\over 4} \sum_{j=0}^{a_1-2} 
\E \left\{ L_{a_1-2-j} L_{j} [ L^o_{a_2} \cdots L^o_{a_q}]^o \right\} +
{(-1)^{a_1+1}\over 4N} \E \left\{ L_{a_1 - 2}^o L_{a_2}^o \cdots
L_{a_q}^o\right\}
$$
$$
+{1+(-1)^{a_i+1} \over 4 N^2}\sum_{i=2}^q 
\E \left\{ L^o_{a_2} \cdots L^o_{a_{i-1}} 
\ a_i \ L_{a_i+a_1-2}L^o_{a_{i+1}} \dots L^o_{a_q}\right\}.
\eqno (3.21)
$$
Comparing this equality with (3.9) and then with (3.10),
we see that $M^{(\eta = -1)}_{2k}$ and  $D^{(q,\eta = -1)}_{2k}$  
are bounded by the elements $B$ and $R^{(q)}$ of recurrent relations
determined 
by equalities  (3.11) and (3.12), respectively. Indeed,
taking into account positivity of $M_{2j}$ and the fact that
$1/4N \le k/4N$ and $1+(-1)^{{a_i}+1}\le 2$,
we obtain inequalities
$$
M_{2k} \le {1\over 4} \sum_{j=0}^{k-1} M_{2j} M_{2k-2-2j} +{1\over 4N} M_{2k-2} +
{1\over 4}\bar D^{(2;\eta = -1)}_{2k-2},
$$
where
$$
\bar D^{(q,\eta = -1)}_{2k} \le \bar D^{(q,\eta = 1)}_{2k}.
$$
Then we conclude that
$$
M^{(\eta = -1)}_{2k} \le 
M^{(\eta = 1)}_{2k} \le B_k \quad {\hbox{and}} \quad
\bar D^{(q,\eta =-1)}_{2k} \le R_{k}^{(q)}.
\eqno (3.22)
$$
This completes the proof of the first part of Theorem 3.2.

\vskip 0.5cm

Now let us turn to the asymptotic expansion of $M_{2k}$
and $D^{(2)}_{2k}$ for fixed $k$ and $N\to\infty$. Regarding (3.21)
with $q=2$, we obtain the following relation
$$
D^{(2)}_{2k} = {1\over 2} \left( D^{(2)} * M\right)_{2k-2} + {1\over 4}
D^{(3)}_{2k-2}+
{(-1)^{a_1 +1}\over 4N}\sum_{a_1 + a_2 = 2k} \E\left\{ L^o_{a_1 -1}
L^o_{a_2}\right\} 
$$
$$ 
+{1+(-1)^{a_2+1}\over 4N^2}\sum_{a_1+a_2= 2k} a_2 \E L_{a_1+a_2
-2}.
$$
Now, introducing variable $r_{k}$ 
$$
D^{(2)}_{2k} = {r_{k}\over N^2} + O(N^{-3})
$$
and taking into account estimates (3.4) and
(3.5), we conclude after simple computations 
that $r$ is determined by recurrent
relations
$$
r_{k} = {1\over 2} \left( r* m\right)_{k-1} + {k^2\over 2}m_{k-1}.
$$
Using relation (5.10),
we can write that
$$
{k^2\over 2} m_{k-1} = {2k(2k-1)\over 2\cdot 4} m_{k-1} + {k\over 4} m_{k-1}=
{1\over 4} \left[{\tau\over (1-\tau)^{3/2}}\right]_k +
{1\over 4} \left[ {\tau \over (1-\tau)^{1/2}}\right]_k.
$$
Then 
$$
r_k = {1\over 4} \left[ {\tau\over (1-\tau)^2} \right]_k + 
{1\over 4}\left[{\tau\over 1-\tau} \right]_k = {k+1\over 4}.
\eqno (3.23)
$$

Now let us consider $1/N$-expansion for $M_{2k}$
$$
M_{2k} = m_k + {1\over N} m_k^{(1)}.
$$
It is easy to see that equality (3.20) together with estimates
(3.5) implies the
following recurrent relation for
$m_k^{(1)}$ 
$$
m_k^{(1)} = {1\over 2} \left( m^{(1)}*m\right)_{k-1} - {m_{k-1}\over 4}.
$$
Then
$$
m^{(1)}_k = - \left[ { \tau f(\tau)\over 4\sqrt{1-\tau} } \right]_k 
= {\delta_{k,0} - (k+1)m_k\over 2}.
\eqno (3.24)
$$
These computations prove the second part of Theorem 3.2.

\vskip 0.5cm
Theorems 3.1 and 3.2
show that there exists essential  difference between GOE and Gaussian anti-symmetric ensemble.
Let us illustrate this by the direct computation of $M_{2}^{(N)}$ for these two
ensembles.

In the case of GOE, we have
$$
{1\over N} \sum_{x,y=1}^N \E H_{xy}^2 = {2\over N^2} \sum _{x<y} \E V_{xy}^2 + 
{1\over N^2} \sum_{x=1}^N \E V_{xx}^2 = {N(N-1)\over 4N^2} + {1\over 2N} =
{1\over 4} + {1\over 4N}. 
$$
This relation reproduce (3.2) with $k=1$.

The first nontrivial moment of anti-symmetric matrices reads as
$$
{1\over N} \sum_{x,y=1}^N \E H_{xy}^2 = {2\over N^2} \sum _{x<y} \E W_{xy}^2 = 
{N(N-1)\over 4N^2} = {1\over 4} - {1\over 4N}
$$
that agrees with (3.24).

Finally, let us point out the difference between GOE and anti-symmetric ensemble
with respect to the first term of the expansion of $D^{(2)}$ given by 
(3.3) and (3.23),
respectively.   This indicates  that
Gaussian Hermitian anti-symmetric ensemble represent a different universality class
 of
the  spectral properties of random matrices  than that of GOE
(see for example, the monograph \cite{M}). 








\section{Gaussian Band random matrices}

Now let us consider the ensemble of Hermitian random matrices given by 
the formula
$$
\left[ H^{(N,b)}\right]_{xy} = h_{xy} \sqrt{U_{xy}}, \quad 
x,y = 1, \dots ,N,
\eqno (4.1)
$$
where $\{h_{xy}, \ x\le y\}$ are determined by (2.1) and (2.2) with $\eta= 0$. 
The elements of non-random matrix $U = U^{(N,b)}$ are determined by
relation
$$
U_{xy} = {1\over b}\  u\left( {x-y\over b}\right),\quad x,y = 1,\dots,N,
$$
where $u(t), \ t\in {\bf R}$ is a positive even piece-wise continuous
function such as
$$
\sup_{t\in {\bf R}} u(t) = u_0 <\infty \quad
{\hbox{and}} \quad \int_{-\infty}^\infty u(t) \ dt = u_1.
$$
Without loss of generality, we can consider $u_0=1$. We assume also that
\mbox{$u(t), t\ge 0 $} is monotone.
If $u(t)$ is given by the indicator function of the interval
$(-1/2,1/2)$, then matrices (4.1) are of the band form.
We keep the term  of band random matrices when regarding the ensemble (4.1)
in the general case.

It is known (see for instance \cite{CG,MPK}) that the moments of $H^{(N,b)}$ converge in the limit of
$1\ll b\ll N$ to the moments of the semicircle law;
$$
M_{2k+1}^{(N,b)} = 0, \quad M_{2k}^{(N,b)} =  \E\left\{{1\over N} \T
\left[H^{(N,b)}\right]^{2k}\right\}
\to m_k(u_1),
\eqno (4.3)
$$
where the numbers $\{m_k(u_1), \ k\ge 0\}$ are given by recurrent
relations
$$
m_k(u_1) = {u_1\over 4} \sum_{j=0}^{k-1} m_{k-1-j}(u_1) \ m_j(u_1), \quad
m_0(u_1)= 1.
$$
The generating function $f_1(\tau) = \sum \tau^k\, m_k(u_1)$ is
related with 
$f(\tau)$ (1.7) by equality
$f_1(\tau) = f(\tau u_1)$ and therefore
$$
m_k(u_1) = u_1^k \, m_k.
$$

\subsection{Main results }

In this section we present non-asymptotic estimate for the moments
of $M_{2k}^{(N,b)}$. This improves proposition (4.3). Let us denote 
$$
\hat u_1 = \hat u_1^{(b)}= {1\over b} + {1\over b} \sum_{l=-\infty}^{+\infty} u\left({l\over
b}\right).
$$
Clearly $\hat u_1 \ge u_1$ and $\h u_1^{(b)} \to u_1$ as $b\to\infty$.

\vskip 0.5cm
\noindent {\bf Theorem 4.1}.

\noindent {\it Given $\alpha> 1/12$, there exists $\theta>0$ such that 
the estimate
$$
M_{2k}^{(N,b)} \le \left( 1 + \alpha\hat u{(k+1)^3\over b^2} \right) m_k(\h u_1),
\eqno (4.4)
$$
where $\h u = \max \{\hat u_1, 1/8\}$, holds for all values of $k,b$ such that
$
{\displaystyle (k+1)^3\over \displaystyle b^2} \le \theta
$
and $b \le N$.
}

\vskip 1cm
The proof of this theorem is obtained by using the method described in section 2.
We consider mathematical expectations of variables $L_{2k}(x) = (H^{2k})_{xx}$
and derive recurrent relations for them and related covariance variables. 
Certainly, these relations are of more complicated structure than those
derived for GUE in section 2.
However, regarding the estimates for $M_{2k} = \E L_{2k}(x)$
by auxiliary numbers $\b B_k$, one can observe that equalities  for 
$\b B_k$ and related numbers $\b R^{(q)}_k$ is almost the same as
the system (2.22)-(2.23) 
derived for GUE.
This allows us to say that the system (2.22)-(2.23) plays an important role
in random matrix theory and is of somewhat canonical character.
The estimates for the moments $M_{2k}^{(N,b)}$ follow immediately.

\subsection{Moment relations and estimates}

In what follows, we omit superscripts $(N,b)$ when no confusion can arise.
It follows from integration by parts formula (5.8) that
$$
\E \left\{ H_{xy} \ (H^l)_{yx}\right\} = 
{1\over 4} \cdot U_{xy} \ \sum_{j=0}^{l-1} 
\E \left\{ (H^j)_{yy} \ (H^{l-j})_{xx} \right\}.
\eqno (4.5)
$$
Then, regarding  $L_{k} (x) = (H^{k})_{xx}$, we obtain 
equality
$$
\E L_{2k}(x) = {1\over 4} \sum_{j = 0}^{2k-2} 
\E \left\{ L_{2k-2-j}(x) \ L_{j}[x] \right\},
$$
where we denoted 
$$
L_j[x] = {1\over b} \sum_{y =1}^N \ u\left( {x-y \over b
}\right)(H^j)_{yy} \ .
$$
Introducing variables $M_{k} (x) = \E L_{k}(x) $ and
$M_{k}[x] = \E L_k[x]$, we obtain equality
$$
M_{2k}(x) = {1\over 4} \sum_{j=0}^{k-1} M_{2k-2-2j}(x) \ M_{2j}[x] +
{1\over 4} D_{2k-2}^{(2)} (x,[x]),
\eqno (4.6)
$$
where we denoted
$$
D_{2k-2}^{(2)}(x,[x]) = \sum_{a_1+a_2=2k-2} 
\E \left\{ L^o_{a_1}(x) \ L^o_{a_2}[x]\right\}.
\eqno (4.7)
$$
In (4.6) we have used obvious equality $M_{2k+1}(x) = 0$.

To get the estimates to the terms standing on the right-hand
sides of (4.6) and (4.7), we need to consider more general expressions
than $M$ and $D$ introduced above.
Let us consider the following variables

$$
M_{2k}^{(\pi_r,\b y_r)}( x) = 
\E \left\{ \left( H^{p_1} \P_{y_1} H^{p_2} \cdots \P_{y_r}
H^{p_{r+1}}\right)_{xx}\right\}, 
\eqno (4.8)
$$
where we denoted $\pi_r = (p_1,p_2,\dots, p_{r+1})$ with
$\sum_{i=1}^{r+1} p_i = 2k$, the vector $\b y_r = (y_1,\dots, y_r)$
and $\P_y$ denotes the diagonal matrix
$$
(\P_y)_{st} = \delta_{st} \ U\left( {t-y\over b}\right), \quad s,t =
1,\dots, N.
$$
One can associate the right-hand side of (4.8) with $2k$ white balls
separated into $r+1$ groups by $r$ black balls.

The second variable we need is
$$
D_{a_1,a_2, \dots, a_q}^{(q, \pi_r(\a_q), \b y_r)}(\b x_q) =
\E \{\ 
\underbrace{ L^o_{a_1}(x_1) L^o_{a_2}[x_2] \cdots L^o_{a_q}[x_q]}
_{\pi_r(\b y_r)}
\ \},
\eqno (4.9)
$$
where $\a_q = (a_1, \dots, a_q)$ and  $\bar x_q = (x_1, \dots, x_q)$. 
We also denote $\vert \a_q\vert = \sum_{i=1}^q a_i$.
So, we have the set of $\vert \a_q\vert $ white balls 
separated into $q$ boxes by $q-1$ walls.

The
brace under the last product means that the set
 $\{a_1| a_2| \cdots |a_q\}$ of walls and white balls  
is separated into $r+1$ groups  by $r$ black balls. The places where the
black balls are inserted depend on the vector $ \a_q$.

Let use derive recurrent relations for (4.8) and (4.9).
These relations resemble very much those obtained in section 2. 
First, we write identity
$$
M_{2k}^{(\pi_r,\b y_r)}( x) = \sum_{s=1}^N 
\E \left\{ H_{xs} \left( H^{p_1-1} \P_{y_1} H^{p_2} \cdots \P_{y_r}
H^{p_{r+1}}\right)_{sx}\right\}, 
$$
and apply the integration by parts formula (4.5). We obtain equality
$$
M_{2k}^{(\pi_r,\b y_r)}( x)   = \sum_{a_1+a_2=2k-2}\E \{\,  
\underbrace{L_{a_1}(x) L_{a_2}[x]}_{\pi'_r(\b y_r,\a_2)} \ \}.
$$
In this relation the partition $\pi'$ 
is different from the original $\pi$
from the left-hand side. It is not difficult to see that $\pi'$
depends  on particular values of $a_1$ and $a_2$, i.e. on the vector
$(a_1,a_2)$. Returning to the denotation $M = \E\{L\}$, we obtain the
first main relation
$$
M_{2k}^{(\pi_r,\b y_r)}( x)   = \sum_{a_1+a_2=2k-2} 
\underbrace{M_{a_1}(x)\,  M_{a_2}[x]}_{\pi'(y_r,\a_2)}
+ \sum_{a_1+a_2=2k-2} D_{a_1,a_2}^{(2,\pi'(y_r,\a_2))}(x,[x]).
\eqno (4.10)
$$

Let us consider 
$$
D_{a_1,a_2, \dots, a_q}^{(q, \pi_r(\a_q), \b y_r)}(\b x_q) =
\sum_{s=1}^N
\E \{\ 
\underbrace{ H_{x_1s}(H^{a_1-1})_{sx_1} \left[L^o_{a_2}[x_2] \cdots
L^o_{a_q}[x_q]\right]^o} _{\pi_r(\b y_r, \a_q)}
\ \}
$$
and apply (4.5) to the last mathematical expectation. We get 
$$
D_{a_1,a_2, \dots, a_q}^{(q, \pi_r(\a_q), \b y_r)}(\b x_q) =
{1\over 4} \sum_{a'=0}^{a_1-2} \E \{ \ 
\underbrace{
L_{a_1-2-a'}(x_1)L_{a'}[x_1] 
\left[L^o_{a_2}[x_2]\cdots L^o_{a_q}[x_q]\right]^o}
_{\pi'_r(\b y_r, \a'_{q+1})}
\ \}
$$
$$
+{1\over 4b^2} \sum_{i=2}^q \sum_{j=0}^{a_i-1}
\E \{ \
\underbrace{\left(H^j \P_{x_i}H^{a_i-1-j} \P_{x_1} H^{a_1-1}
\right)_{x_1x_1} 
L^o_{a_2}[x_2]\cdots \times_i \cdots
 L^o_{a_{q}}[x_{q}]}_
{\pi''_{r+2}(\b y'_{r+2}, \a''_{q+1}(i))}
\ \}.
\eqno (4.11)
$$
In these expressions, $\pi'$ and $\pi''$ designate  partitions 
different from $\pi$; 
they depend on the vectors $\a'_{q+1} = (a_1-2-a',a',a_2,\dots,a_q)$
and $$
\a''_{q+1} (i) = (j,a_i-1-j,a_1-1, a_2,\dots, a_{i-1},a_{i-1},\dots
,a_q),
$$
respectively;
also $\b y'_{r+2} = (x_i,x_1,y_1,y_2,\dots,y_r)$. The notation 
$\times_i$ in the last product of (4.11) means that the factor 
$L_{a_i}$ is absent there. Repeating the computations of section 2,
we arrive at the second main relation
$$
D_{a_1,a_2, \dots, a_q}^{(q, \pi_r(\a_q), \b y_r)}(\b x_q) =
\sum_{l=1}^6 T_l,
\eqno (4.12)
$$
where
$$
T_1 = {1\over 4} \sum_{a'=0}^{a_1-2} \ 
\underbrace{M_{a_1-2-a'}(x_1)\  D_{a',a_2,\dots,a_q}^{(q)}}
_{\pi'_r(\b y_r, \a'_{q+1})}([x_1],[x_2], \dots, [x_q]);
$$
$$
T_2 = 
{1\over 4} \sum_{a'=0}^{a_1-2} \ 
\underbrace{M_{a_1-2-a'}[x_1]\  D_{a',a_2,\dots,a_q}^{(q)}}
_{\pi'_r(\b y_r, \a'_{q+1})}(x_1,[x_2], \dots, [x_q]);
$$
$$
T_3 = 
{1\over 4} \sum_{a'=0}^{a_1-2}
D_{a_1-2-a',a',a_2,\dots,a_q}^{(q+1,\pi'_r(\b y_r, \a'_{q+1}))}
(x_1,[x_1],[x_2], \dots, [x_q]);
$$
$$
T_4 = 
-{1\over 4} \sum_{a'=0}^{a_1-2}\ 
\underbrace{
D_{a_1-2-a',a'}^{(2)}(x_1,[x_1]) \ D_{a_2,\dots, a_q}^{(q-1)}}_
{\pi'_r(\b y_r, \a'_{q+1})}
([x_2],\dots, [x_q]);
$$
$$
T_5 = {1\over 4b^2} \sum_{i=2}^{q}
\sum_{j=0}^{a_i-1} 
\
\underbrace{M_{a_1+a_i - 2}(x_1)\
D^{(q-2)}_{a_2,\dots,a_{i-1},a_{i+1},\dots,a_q}}_
{ \pi_{r+2}''(\b y'_{r+2}, \a''_{q+2}(i))} 
\ ([x_2],
\dots,[x_{i-1}],[x_{i+1}],\dots, [x_q]);
$$
and finally
$$
T_6 = 
{1\over 4b^2} 
\sum_{i=2}^{q}
\sum_{j=0}^{a_i-1} D_{a_1+a_i-2,a_2,\dots, a_{i-1},a_{i+1},\dots, a_q}
^{(q-1,  \pi''_{r+2}(\b y'_{r+2},\a''_{q}(i)))}
(x_1, [x_2], \dots,[x_{i-1}],[x_{i+1}],\dots,[x_{q}]).
$$

Now let us introduce auxiliary numbers $ \{\h B^{(N,b)}_k,\  k\ge 0\}$
and 
$$
\h R^{(q; N,b)}_{\a_q} = \h R_{a_1,\dots, a_q}^{(q; N,b)}, 
\quad {\hbox{for}}\quad q \ge 0 \ \hbox{and} \  a_i \ge 0,
$$
determined for all integer $k,q $ and $a_i$ by the following recurrent
relations (in $\h B$ and $\h R$, we omit superscripts
$N$ and
$b$).
Regarding $\{\h B\}$, we set 
$
\h B_0 = 1
$ and determine $\hat B_k$ by relation
$$
\h B_k = {\h u_1\over 4} \sum_{j=0}^{k-1} \h B_{k-1-j} \h B_j + {1\over
4} 
\sum_{a_1+a_2 = 2k-2} \h R_{a_1,a_2}^{(2)}, \quad k\ge 1.
\eqno (4.13)
$$

Regarding $\{ \h R\}$, we set $\h R^{(0)}=1$ and $\h R^{(1)}_{a} = 0$.
We also assume that $\h R^{(q)}_{\a_q} =0$ when either 
$q>\vert \a_q\vert$ or one of the variables $a_i$ is equal to zero.
The recurrent relation for $\h R$ is
$$
\h R^{(q)}_{a_1,\dots, a_q} = 
{\h u_1\over 2} \sum_{j=0}^{a_1-2-j} \h B_{a_1-2-j} 
\h R^{(q)}_{j,a_2,\dots, a_q}  
$$
$$
+{1\over 4}\sum_{j=0}^{a_1-2-j} \h R^{(q+1)}_{j,a_1-2-j,a_2,\dots,
a_q}+
{1\over 4} \sum_{j=0}^{a_1-2-j} 
\h R^{(2)}_{a_1-2-j,j}
\h R^{(q-1)}_{a_2,\dots, a_q}
$$
$$
+
{\h u_1\over 4b^2} \sum_{i=2}^q a_i \h B_{a_1+a_I-2} 
\h R^{(q-2)}_{a_2,\dots, a_{i-1},a_{i+1},\dots,a_q}+
{1\over 4b^2} \sum_{i=2}^q a_i 
\h R^{(q-1)}_{a_2, \dots, a_{i-1}, a_{i+1}, \dots, a_q}.
\eqno (4.14)
$$
Existence and uniqueness of the  numbers $\h B$ and $\h R$ follow from
the triangular scheme described above in section 2.

Using the  triangular scheme of section 2, 
it is easy to deduce from relations (4.10) and (4.11) that 
$$
\sup_{x,\b y_r}M_{2k}^{(\pi_r,\b y_r)}(x) \le \h B_k
\eqno (4.15)
$$
and 
$$
\sup_{\b x_q, \b y_r} \vert 
D_{a_1,a_2,\dots, a_q}^{(q, \pi_r(\a_q),\b y_r)}
\left(x_1, [x_2], \dots,[x_q]\right)\vert 
\le \h  R^{(q)}_{a_1,a_2, \dots, a_q}.
\eqno (4.16)
$$
Let us note that when regarding (4.15) with $k=0$, we have used 
the property of  $u$ (4.2)
$$
M_0^{(\pi_r, \b y_r)}(x) = \prod_{i=1}^r u \left( {x-y_i\over b}\right)
\le u_0^r \le 1.
$$

Now, let us introduce two more auxiliary sets of numbers $\b B_k$ and
$
\b R{(q)}_{k}$. We  determine them  by relations
$$
\b B_k = {\h u_1\over 4} \sum_{j=0}^{k-1} \b B_{k-1-j} \b B_j + 
{1\over 4}\b R_{k-1}^{(2)},\quad \bar B_0=1,
\eqno (4.17)
$$
and
$$
\b R^{(q)}_{k} = {\h u_1\over 2} 
\sum_{j=0}^{k-1} \b R^{(q)}_{k-1-j} \b B_j + 
{\h u_1(q-1)\over 4b^2}\sum_{j=0}^{k-1} \b R^{(q-2)}_{k-1-j} 
{(2j+2)(2j+1)\over 2} \b B_j 
$$
$$
+{1\over 4} \b R^{(q+1)}_{k-1} + {1\over 4} \sum_{j=0}^{k-1}
\b R^{(2)}_{k-1-j} \b R^{(q-1)}_{j} + {2k^2(q-1)\over 4b^2}
\b R^{(q-1)}_{k-1}.
\eqno (4.18)
$$

It is clear that 
$$
\h B_k \le \b B_k
\qquad {\hbox{and}} \quad
\sum_{a_1+ \dots a_q = 2k} \h R_{a_1, \dots, a_q}^{(q)} \le \b
R_{k}^{(q)}.
\eqno (4.19)
$$
The main technical result of this section is as follows.


\vskip 0.5cm
\noindent {\bf Proposition 4.1.} 

\noindent {\it Let $\h u = \max \{\h u_1, 1/8 \}$. Given $A>1/16$, there exists $\th>0$
such that the estimate
$$
\b B_k \le \left[ f_1(\tau) + {A\h u\over b^2} {\tau ^2\over (1 - \tau \h u_1)^{5/2}}\right]_k
\eqno (4.20)
$$
holds for all values of $k\le k_0$, where $k_0$ verifies condition 
$k_0^3 \le \th b^2$.
Also there exists $C$ 
$$
{1\over 24} < C < \max \{{3A\over 2},  4!\}
$$
such that inequalities
$$
\b R^{(2s)}_{k} \le 
C {\h u^s(3s)!\over b^{2s}} \left[ {\tau\over
(1-\tau \h u_1)^{2s}}
\right]_{k}
\eqno (4.21a)
$$
and 
$$
R^{(2s+1)}_{k} \le 
C{\h u^{s+1}(3s+3)!\over b^{2s+2}} 
\left[ {\tau \over (1-\tau \h u_1)^{2s+1}} \right]_{k},
\eqno (4.21b)
$$
hold for all values of $k$ and $s$ such that 
$$
2k+ q \le 2k_0
$$
with $q= 2s$ and $q=2s+1$, respectively. 
}

\vskip 0.5cm

The proof of this proposition can be obtained by repeating 
the proof of Proposition 2.1 with obvious changes. The only difference 
is related with the presence of the factors $\h u_1$ in (4.17) and (4.18). This
implies corresponding changes in 
the generating functions used in estimates (4.20) and (4.21).
Also, the conditions for $A$ (2.29) and $C$ (2.38), (2.43) are replaced by conditions
$$
A > {3C\over 2} + {A^2 \h u_1\over 16} \th,
$$
$$
C > { \delta_{s,1} + C (1-\delta_{s,1})\over 24} 
+
2 \theta \left( 1 + 10 \hat u C(1+C) + 2\h u_1 AC\right)
$$
and
$$
179 \h u > 20 + 3\h u C + 18 \th \h u \h u_1 A.
$$
The last inequality forces us to 
use $\h u$ instead of $\h u_1$ in the proof. 
Otherwise, we should assume
that  $\hat u_1 > 1/8$. We believe this condition is technical and can be avoided.

\subsection{Spectral norm of band random matrices}

Using this result, we can estimate the lower bound for $b$ to have the spectral norm
of $\Vert H^{(N,b)}\Vert = \lambda_{\max}^{(N,b)}$ bounded.

\vskip 0.5cm
\noindent {\bf Theorem 4.2}
If $1\ll (\log N)^{3/2} \ll b$, then $\l_{\max}^{(N,b)} \to \sqrt{u_1}$\, with
probability 1. 

\vskip 0.5cm
\noindent {\it Proof.} Using the standard inequality
$$
P\left\{ \l_{\max}^{(N,b)} > \sqrt{u_1}(1+ \vep)\right\} \le N{ M_{2k}^{(N,b)}\over u_1^k
(1+\vep)^{2k}},
$$
we deduce from (4.4) estimate
$$
P\left\{ \l_{\max}^{(N,b)} > \sqrt{u_1} (1+ \vep) \right\} \le 
N { \left( 1 + \alpha \hat u {\displaystyle (k+1)^2\over \displaystyle b^2}\right)^k\over
u_1^k (1+\vep)^{2k}Ê} \hat u_1^k
\eqno (4.22)
$$
that holds for all $k+1 \le \theta ^{1/3} b^{2/3}$, where $\theta $ is as in Theorem 4.1.
In (4.22), we have used inequalities $m_k(\hat u_1) \le \h u_1^k\,  m_{2k}$ and 
$m_{2k} \le 1$. 

Assuming that $b = \phi_N (\log N)^{3/2}$, where $\phi_N \to \infty$ as $N\to \infty$,
and taking $k+1 = t\theta^{1/3}  b^{2/3}$, $0<t\le 1$,
we  obtain the estimate
$$
P\left\{ \l_{\max}^{(N,b)} > \sqrt{u_1} (1+ \vep) \right\} \le 
N \exp\left\{- 2t \theta ^{1/3} b^{2/3} \log (1+\vep) + 2\alpha \hat u t^3\right\} \cdot
\left( {\hat u_1\over u_1} \right)^k.
\eqno (4.23)
$$
Using relation $\hat u_1 = u_1(1 + 1/b)$, we easily deduce from 
(4.23) that
$$
P\left\{ \l_{\max}^{(N,b)} > \sqrt{u_1} (1+ \vep) \right\} \le
 N^{1- C \log (1+\vep) \phi_N^{2/3}}
$$
with some positive $C $. 
Then corresponding series of probability converges and the Borel-Cantelli lemma
implies convergence of $\lambda_{\max}^{(N,b)}$ to $\sqrt{u_1}$.
Theorem 4.2 is proved.

\vskip 0.5cm

Let us complete this subsection with the following remark.
If one optimizes the right-hand side of (4.23), 
one can see that the choice of 
$t=t_0 = b^{1/3} \sqrt{\log (1+\vep)} (\alpha \h u)^{-1/2} \theta ^{-1/3}$ gives the best
possible estimate in the form
$$
N \exp\{ - b{1\over \sqrt {2\alpha \h u}} (\log (1+\vep))^{3/2}\}.
$$
Once this estimate shown, convergence $\lambda_{\max}^{(N,b)}\to \sqrt {u_1}$
would be true under condition that  $b= O(\log N)$. 
However, one cannot use the optimal value of $t_0$ mentioned above because this choice
makes
$k$ to be
$k = O(b)$. This asymptotic regime is out of reach for the method of this paper.

\section{Auxiliary relations}

\subsection{Integration by parts for complex random variables}

Let us consider matrices $H$ 
 with elements  $H_{xy} = v_{xy}+ \i w_{xy}$,
 where the family 
 $\{v_{xy}, w_{xy}, \ 1\le  x\le y\le N\}$ is given by  jointly independent Gaussian random variables 
with zero mean value.
We denote 
$$
\E v^2_{xy} = \xi_{xy}, \quad \E w^2_{xy} = \zeta_{xy}.
$$
Let us assume that $x<y$.
Then integration by parts formula says that
$$
\E H_{xy} (H^l)_{st} = 
\xi_{xy} \E \left\{ {\partial (H^l)_{st}\over \partial
v_{xy} } \right\}
+
\i \zeta_{xy} \E \left\{ {\partial (H^l)_{st}\over \partial
w_{xy} } \right\}
\eqno (5.1)
$$
It is easy to see that
$$
 {\partial (H^l)_{st}\over \partial
v_{xy} } = \sum_{j=1}^l \sum_{s',t'=1}^N H^{j-1}_{ss'}\cdot  
{\partial H_{s't'}\over \partial
v_{xy} } \cdot H^{l-j}_{vt} 
= \sum_{j=1}^l \left[ H^{j-1}_{sx} H^{l-j}_{yt} +
H^{j-1}_{sy} H^{l-j}_{xt}\right].
\eqno (5.2)
$$
Similarly 
$$
 {\partial (H^l)_{st}\over \partial
w_{xy} } = 
\i \sum_{j=1}^l \left[ H^{j-1}_{sx} H^{l-j}_{yt} -
H^{j-1}_{sy} H^{l-j}_{xt}\right].
\eqno (5.3)
$$
Substituting (5.2) and (5.3) into (5.1), we get equality
$$
\E H_{xy} (H^l)_{st} = 
\left( \xi_{xy} - \zeta_{xy}\right) \sum_{j=1}^l
\E\{ H^{j-1}_{sx}
H^{l-j}_{yt}\}
$$
$$+
\left( \xi_{xy} + \zeta_{xy}\right) \sum_{j=1}^l
\E\{ H^{j-1}_{sy}
H^{l-j}_{xt}\}, \quad x<y.
\eqno (5.4)
$$
It is not hard to check that the same relation is true 
when $x>y$.
Also
$$
\E H_{xx} (H^l)_{st} = \xi_{xx} \sum_{j=1}^l \E\{H^{j-1}_{sx}
H^{l-j}_{xt}\}.
\eqno (5.5)
$$

\subsubsection{Gaussian Ensembles $\{H^{(\eta)}\}$}

Regarding  formulas (2.1)-(2.3), we see that
$$
 v_{xy} = {V_{xy}\over \sqrt N}, \quad w_{xy} = {W_{xy}\over \sqrt N}
 $$
and 
$$
\xi_{xy} + \zeta_{xy} = {1+ \delta_{xy}\eta\over 4N}, \quad 
\xi_{xy} - \zeta_{xy} = {\eta + \delta_{xy}\over 4N}.
$$
Regarding the sum of (5.5) with doubled (5.4), 
we obtain relation valid for all values of $x$ and $y$
$$
\E H_{xy} (H^l)_{st} = {1\over 4N} \sum_{j=1}^{l} 
\E\{H^{j-1}_{sy}
H^{l-j}_{xt}\} +
{\eta\over 4N} \sum_{j=1}^{l}  
\E\{ H^{j-1}_{sx} 
H^{l-j}_{yt}\}.
\eqno (5.6)
$$
Let us mention two useful formulas that follow from (5.6); these are
$$
\E \T (H^{l+1}) = {1\over 4N} \sum_{j=1}^l\E \left\{ \T H^{j-1} \T
H^{l-j}\right\} 
$$
$$
+{\eta\over 4N}\sum_{j=1}^l \sum_{x,y=1}^N\E \left\{ (H^{j-1})_{yx} (H^{l-j})_{yx}
\right\}
\eqno (5.7a)
$$
and
$$
\E  H_{xy} \T H^l = \E \left\{H_{xy} \sum_{s=1}^N
(H^l)_{ss}\right\} =  {l\over 4N} \E H^{l-1}_{xy} +
{\eta l\over 4N}
\E H^{l-1}_{yx}.
\eqno (5.7b)
$$

\subsubsection{Band Random Matrices}

Using (5.4) and (5.5) in the case of matrices (4.1), we see that
$$
\xi_{xy} = {1+\delta_{xy}\over 8} U_{xy}, \quad \zeta_{xy} = {1-\delta_{xy}\over 8} U_{xy}.
$$
Then (5.4) and (5.5) imply equality
$$
\E H_{xy} (H^l)_{st} = {U_{xy}\over 4} \sum_{j=1}^{l} 
\E\{H^{j-1}_{sy}
H^{l-j}_{xt}\} .
\eqno (5.8)
$$
Regarding this relation, one can easily obtain analogs of formulas (5.7a) and (5.7b).

\subsection{Derivation of Equality (2.18)} 

We consider the case of Hermitian matrices $\eta=0$ only. Regarding (2.17), 
we can write that
$$
\E\{ L^o_{a_1}\cdots L^o_{a_q}\} = \E \{ L_{a_1} Q^o\},
$$
where $Q = L^o_{a_2}\cdots L^o_{a_q}$. Using integration by parts formula,
we obtain as in (5.1) 
that
$$
\E \{ H_{xy} (H^{a_1-1})_{yx} Q^o\} = 
\xi_{xy} \E \left\{ { \partial H^{a_1-1}_{yx}Q^o\over \partial v_{xy} } \right\} 
+ \i \zeta_{xy} \E \left\{ { \partial H^{a_1-1}_{yx}Q^o\over \partial w_{xy} } \right\} .
$$
Obviously,
$$
 { \partial H^{a_1-1}_{yx}Q^o\over \partial v_{xy} } = 
 \sum_{j=1}^{a_1-1}\E \left\{\left[H^j_{yx}H^{a_1-1-j}_{yx} + H^j_{yy}H^{a_1-1-j}_{xx}\right] Q^o\right\}
 + H^{a_1-1}_{yx} {\partial Q^o\over \partial v_{xy}}.
 $$
It is clear that
$$
 {\partial Q^o\over \partial v_{xy}} =  {\partial Q\over \partial v_{xy}}=
 \sum_{i=2}^q L^o_{a_2} \cdots L^o_{a_{i-1}}{\partial L^o_{a_i}\over \partial v_{xy}} L^o_{a_{i+1}}\dots
 L^o_{a_q}
  $$
and 
$$
{\partial L_{a_i}\over \partial v_{xy} } =
{1\over N} \sum_{t=1}^N \sum_{j=1}^{a_i} \left[ H^{j-1}_{tx} H^{a_i-j}_{yt}
+ H^{j-1}_{ty} H^{a_i-j}_{xt}\right] = {a_i\over N} \left[H^{a_i-1}_{xy} + H^{a_i-1}_{yx}\right].
$$

Also we have
$$
 { \partial H^{a_1-1}_{yx}Q^o\over \partial w_{xy} } = 
\i  \sum_{j=1}^{a_1-1}\E \left\{\left[H^{j-1}_{yx}H^{a_1-1-j}_{yx} - H^{j-1}_{yy}H^{a_1-1-j}_{xx}\right] Q^o\right\}
 + H^{a_1-1}_{yx} {\partial Q^o\over \partial w_{xy}}.
 $$
It is clear that
$$
 {\partial Q^o\over \partial w_{xy}} =  {\partial Q\over \partial w_{xy}}=
 \sum_{i=2}^q L^o_{a_2}L^o_{a_{i-1}} \cdots {\partial L^o_{a_i}\over \partial w_{xy}} L^o_{a_{i+1}}\dots
 L^o_{a_q}
  $$
and 
$$
{\partial L_{a_i}\over \partial w_{xy} } =
{\i \over N} \sum_{t=1}^N \sum_{j=1}^{a_i} \left[ H^{j-1}_{tx} H^{a_i-j}_{yt}
- H^{j-1}_{ty} H^{a_i-j}_{xt}\right] = {\i a_i\over N} \left[H^{a_i-1}_{yx}-H^{a_i-1}_{xy}
\right].
$$
Gathering these terms, we finally obtain that
$$
\E \{ H_{xy} (H^{a_1-1})_{yx} Q^o\} = 
{1\over 4N} \sum_{j=1}^{a_1-1} \E \left\{ H^{j-1}_{yy} H^{a_1-1-j}_{xx} Q^o\right\}
$$
$$
+{1\over 4N^2} \sum_{i=2}^{q} a_i \E \left\{ H^{a_1-1}_{yx} 
 L^o_{a_2} \cdots L^o_{a_{i-1}} H^{a_i-1}_{xy} L^o_{a_{i+1}}\dots
 L^o_{a_q}\right\}.
 $$
 Now (2.18) easily follows.

\subsection{Catalan numbers and related identities}

In 
the proofs, we have used the following identity
for any integer $r\ge 1$,
$$
\left[ {1\over (1-\tau)^{r +1/2}} \right]_k = r \
{ { {2k + 2r}\choose{2k}}  \over {k+r\choose k+1}} \ m_k \ , 
\eqno (5.9)
$$ 
or in equivalent form,
$$
\left[ {1\over (1-\tau)^{r +1/2}} \right]_k =
{1\over 2^{2k} \ k! }\cdot {(2k+2r)!\over (2r)!}\cdot {r! \over (k+r)!}.
\eqno (5.10)
$$
Two particular cases are important:
$$
{(2k+2)(2k+1)\over 2 } m_k = \left[ {1\over (1-\tau)^{3/2}}\right]_k.
\eqno (5.11)
$$
and
$$
{(2k+1)(2k+2)(2k+3)  \over 3!} m_k  = 
\left[{1\over (1-\tau)^{5/2}}\right]_k.
\eqno (5.12)
$$
We also use  equality
$$
\left[ {1 \over (1-\tau)^{l+1}} \right]_k = {(k+1) \cdots (k+l)\over l!} 
= {(k+l)!\over k!\  l!}. 
\eqno (5.13)
$$

\vskip 0.5cm
{\bf Acknowledgments.} The author is grateful to Prof. M. Ledoux for the constant
interest to this work and to Prof. A. Rouault for
numerous remarks and comments. The author also thanks
the anonymous referee for the careful reading of the manuscript and for  a number 
of 
corrections and useful suggestions
that improve the presentation.
This work was partially supported by the "Fonds National de la Science
(France)" via the ACI program "Nouvelles Interfaces des
Math\'ematiques", project MALCOM n$^\circ$205.

\end{document}